\documentclass[sigconf]{acmart} 

\usepackage{arydshln}
\usepackage{textcomp}

\makeatletter
\def\adl@drawiv#1#2#3{%
        \hskip.5\tabcolsep
        \xleaders#3{#2.5\@tempdimb #1{1}#2.5\@tempdimb}%
                 #2\z@ plus1fil minus1fil\relax 
        \hskip.5\tabcolsep}
\newcommand{\cdashlinelr}[1]{%
  \noalign{\vskip\aboverulesep
          \global\let\@dashdrawstore\adl@draw
          \global\let\adl@draw\adl@drawiv}
  \cdashline{#1}
  \noalign{\global\let\adl@draw\@dashdrawstore
          \vskip\belowrulesep}}
\makeatother

\newif{\ifhidecomments}
 \hidecommentsfalse 
\ifhidecomments 
    \newcommand{\yifan}[1]{} 
    \newcommand{\janice}[1]{} 
    \newcommand{\help}[1]{} 
\else 
    \newcommand{\janice}[1]{\textbf{\sffamily{\textcolor{purple}{[Janice: #1 ]}}}} 
    \newcommand{\yifan}[1]{\textbf{\sffamily{\textcolor{cyan}{[Yifan: #1 ]}}}} 
    \newcommand{\help}[1]{\textbf{\sffamily{\textcolor{orange}{[NEED HELP: #1 ]}}}}
\fi

\AtBeginDocument{%
  \providecommand\BibTeX{{%
    \normalfont B\kern-0.5em{\scshape i\kern-0.25em b}\kern-0.8em\TeX}}}

\copyrightyear{2024}
\acmYear{2024}
\setcopyright{rightsretained}
\acmConference[L@S '24]{Proceedings of the Eleventh ACM Conference on Learning @ Scale}{July 18--20, 2024}{Atlanta, GA, USA}
\acmBooktitle{Proceedings of the Eleventh ACM Conference on Learning @ Scale (L@S '24), July 18--20, 2024, Atlanta, GA, USA}
\acmDOI{10.1145/3657604.3662036}
\acmISBN{979-8-4007-0633-2/24/07}

\begin{document}

\title{Evaluating the Effectiveness of LLMs in Introductory Computer Science Education: A Semester-Long Field Study}
%

\author{Wenhan Lyu}  
\affiliation{
    \institution{William \& Mary}
    \city{Williamsburg}
    \state{VA}
    \country{USA}} 
\email{wlyu@wm.edu}
\orcid{0009-0004-9129-8689} 

\author{Yimeng Wang}  
\affiliation{
    \institution{William \& Mary}
    \city{Williamsburg}
    \state{VA}
    \country{USA}} 
\email{ywang139@wm.edu}
\orcid{0009-0005-0699-4581} 

\author{Tingting (Rachel) Chung}  
\affiliation{
    \institution{William \& Mary}
    \city{Williamsburg}
    \state{VA}
    \country{USA}}
\email{rachel.chung@mason.wm.edu}
\orcid{0000-0002-0250-4873}

\author{Yifan Sun}  
\affiliation{
    \institution{William \& Mary}
    \city{Williamsburg}
    \state{VA}
    \country{USA}}
\email{ysun25@wm.edu}
\orcid{0000-0003-3532-6521}

\author{Yixuan Zhang}  
 \affiliation{
    \institution{William \& Mary}
    \city{Williamsburg}
    \state{VA}
    \country{USA}} 
\email{yzhang104@wm.edu}
\orcid{0000-0002-7412-4669}

\renewcommand{\shortauthors}{Wenhan Lyu, Yimeng Wang, Tingting (Rachel) Chung, Yifan Sun, \& Yixuan Zhang}

\begin{abstract} 
The integration of AI assistants, especially through the development of Large Language Models (LLMs), into computer science education has sparked significant debate, highlighting both their potential to augment student learning and the risks associated with their misuse. An emerging body of work has looked into using LLMs in education, primarily focusing on evaluating the performance of existing models or conducting short-term human subject studies. However, very little work has examined the impacts of LLM-powered assistants on students in entry-level programming courses, particularly in real-world contexts and over extended periods. To address this research gap, we conducted a semester-long, between-subjects study with 50 students using CodeTutor, an LLM-powered assistant developed by our research team. Our study results show that students who used CodeTutor (the ``CodeTutor group'' as the experimental group) achieved statistically significant improvements in their final scores compared to peers who did not use the tool (the ``control group''). Within the CodeTutor group, those without prior experience with LLM-powered tools demonstrated significantly greater performance gain than their counterparts. We also found that students expressed positive feedback regarding CodeTutor's capability to comprehend their queries and assist in learning programming language syntax. However, they had concerns about CodeTutor's limited role in developing critical thinking skills. Over the course of the semester, students' agreement with CodeTutor's suggestions decreased, with a growing preference for support from traditional human teaching assistants. Our findings also show that students turned to CodeTutor for different tasks, including programming task completion, syntax comprehension, and debugging, particularly seeking help for programming assignments. Our analysis further reveals that the quality of user prompts was significantly correlated with CodeTutor's response effectiveness. Building upon these results, we discuss the implications of our findings for the need to integrate Generative AI literacy into curricula to foster critical thinking skills, and turn to examining the temporal dynamics of user engagement with LLM-powered tools. We further discuss the discrepancy between the anticipated functions of tools and students' actual capabilities, which sheds light on the need for tailored strategies to improve educational outcomes.
\end{abstract}




\maketitle

\section{Introduction} 
Recent advancements in Generative AI and Large Language Models (LLMs), exemplified by GitHub Copilot~\cite{GitHubCopilot} and ChatGPT~\cite{ChatGPT}, have demonstrated their capacity to tackle complex problems with human-like proficiency. These innovations raise significant concerns within the educational domain, particularly as students might misuse these tools, thereby compromising the quality of education and breaching academic integrity norms~\cite{perkins2023detection}. Specifically, entry-level computer science education is directly affected by the progress in LLMs~\cite{zhou2024teachers}. LLMs' capability in handling programming tasks means they can complete many assignments typically given in introductory courses, thus becoming highly appealing to students looking for easy solutions.

Despite these challenges, LLM-powered tools offer great opportunities to enrich computer science education~\cite{kumar2023impact}. When used ethically and appropriately, they can serve as powerful educational resources. For instance, LLMs can provide students instant feedback on their coding assignments or generate diverse examples of code that help demonstrate programming concepts~\cite{pankiewicz2023large}. Moreover, as Generative AIs are becoming popular in production environments, familiarizing students with these technologies is increasingly becoming a crucial aspect of computer science education. 

The unique challenges posed by LLMs stem from the difficulty in detecting the use of AI tools~\cite{wu2023survey, zhou2023synthetic}. Traditional approaches, such as plagiarism detection software, fall short in determining the originality of student submissions~\cite{meyer2023chatgpt}. Given the challenges in identifying LLMs usage and recognizing the potential advantages of these technologies, we consider integrating LLMs into computer science education inevitable. As students have already started using such tools, the impact of LLMs on computer science education remains unknown. Indeed, a growing body of research has begun to explore the application of LLMs within educational settings, primarily focusing on assessing the capabilities of current models with existing datasets or previous assignments from students~\cite{hicke2023chata,mehta2023can}. However, there is still a research gap in understanding how students interact with LLM-powered tools in introductory programming classes, particularly regarding their engagement in genuine learning settings over extended periods. Furthermore, while previous studies have shown individual differences in intelligent tutoring systems~\cite{kulik2016effectiveness}, research into how these differences apply to LLM tools is lacking. Investigating these variations is important for tailoring educational strategies to diverse student needs. In short, understanding these nuanced attitudes of and interactions with LLM-powered tools in CS education over extended periods is crucial for identifying the evolving challenges and opportunities LLMs introduce.

To address the research gap, we asked the following research questions (RQs) in this work: \\
\textbf{RQ1.} Does the integration of LLM-powered tools in introductory programming courses enhance or impair students' learning outcomes, compared to traditional teaching methods? How are individual differences associated with students' learning outcomes using LLM-powered tools? \\
\textbf{RQ2.} What are students' attitudes towards LLM-powered tools, how do they change over time, and which factors might influence these attitudes? \\
\textbf{RQ3.} How do students engage with LLM-powered tools, and how do they respond to their programming needs?

We believe that addressing the following research questions (RQs) is critical for enabling researchers to make informed decisions about incorporating LLMs into their courses and guiding students on the optimal and responsible use of LLM-powered tools. To answer the questions, we conducted a longitudinal, between-subject field study with 50 students over the course of the fall semester from September to December 2023 with a web-based tool we developed called CodeTutor.

The \textbf{contributions} of this work are:
\textbf{1) }We conducted a semester-long longitudinal field study to assess the effectiveness of an LLM-powered tool (CodeTutor) on students' learning outcomes in an introductory programming course. By comparing the performance of students who used CodeTutor against those who did not, our study contributes to new empirical evidence regarding the role of LLM-powered tools in the programming learning experience;
\textbf{2) }We characterized patterns of student engagement with CodeTutor and analyzed the ways in which it can meet students' learning needs. Through the analysis of conversational interactions and feedback loops between students and the tool, we contributed new knowledge regarding how CodeTutor facilitates or impedes learning; and
\textbf{3) }We offered insights and outlined design implications for future research.

\section{Related Work}
\subsection{Intelligent Tutoring Systems}
Using computerized tools for assisting educational purposes is not a new idea. As early as the 1950s, the first concept of using computers to assist learning has already emerged~\cite{nwana1990intelligent}. From where the factor of intelligence had been considered and it had started evolving into \textit{Intelligent Tutoring Systems (ITS)}~\cite{sleeman1982intelligent}. ITS leverages artificial intelligence to provide personalized learning experiences in computer science education, adapting instruction and feedback to individual student needs~\cite{anderson1985intelligent, elsom1984design}. These systems have enhanced student engagement, comprehension, and problem-solving skills by offering tailored support and immediate feedback, similar to one-on-one tutoring~\cite{vanlehn2011relative, demszky2023m}. Research has demonstrated that ITS can significantly improve understanding of complex concepts in programming courses compared to traditional teaching methods, leading to higher student satisfaction due to the personalized learning environment~\cite{corbett1997intelligent, ritter2007cognitive}. The Internet also empowered ITS to offer more interactivity and adaptivity~\cite{brusilovsky1996elm, brusilovsky1998adaptive, butz2006web}, leveraging the path of later boost with natural language processing techniques~\cite{el2008natural, hooshyar2015flowchart}.

However, prior work has shown that as the granularity of tutoring decreases, its effectiveness increases~\cite{vanlehn2011relative}. Significant limitations for ITS include the complexity and cost of building them, the incapability to answer questions and tasks out of their programmed domains, and the difficulty to develop with the purpose of productively used by individuals without expertise~\cite{graesser2018intelligent}. Even though the Generalized Intelligent Framework for Tutoring (GIFT) framework~\cite{sottilare2012generalized} was proposed and evolved for developing ITS for use at scale, those limitations mostly remain unresolved.

\subsection{Large Language Models in CS Education}

The release of ChatGPT and other Generative AI applications brought LLMs into the public view and attracted enormous attention~\cite{achiam2023gpt, sun2024trustllm}. LLMs offer researchers and users the flexibility to employ a single tool across various tasks~\cite{wei2022emergent}, such as medical research~\cite{llm_medicine_1, llm_medicine_2}, finance~\cite{llm_finance_1}, and education~\cite{kasneci2023chatgpt}. Adopting LLM-powered tools in educational settings is facilitated by their broad accessibility and cost-free nature~\cite{zamfirescu2023johnny}. Recent studies have looked into the potential of AI assistants to enhance student learning by helping with students' problem-solving~\cite{ahmed2022synshine, leinonen2023using, phung2023generating} and generating computer science content~\cite{sarsa2022automatic, denny2022robosourcing}. Current research on the use of LLMs in education has primarily looked into their performance and capabilities~\cite{prather2023robots} compared to humans, such as generating code for programming tasks~\cite{leinonen2023comparing, poldrack2023ai}, answering general inquriries~\cite{savelka2023large, phung2023generative}, addressing textbook questions~\cite{jalil2023chatgpt} and exam questions~\cite{dobslaw2023experiences}. 

Despite the growing interest in examining the capabilities of LLMs in education, very few empirical studies have examined the emerging concerns regarding their impact. Therefore, there is an urgent need for research into the long-term effects of LLMs in CS education and the development of strategies to counteract potential negative consequences. One exceptional work was conducted by Liffiton et al.~\cite{liffiton2023codehelp}, who developed a tool called CodeHelp for assisting students with their debug needs in an undergraduate course over 12 weeks. Their follow-up study~\cite{sheese2023patterns} categorized history messages in their tool, and found a positive relationship between tool usage and course performance. However, their study specifically focused on debugging issues and did not compare the outcomes with those achieved through traditional TA methods. 

Furthermore, prior research has demonstrated that individual differences, such as gender, race, and prior experiences with technologies, significantly influence the effectiveness of intelligent tutoring systems~\cite{kulik2016effectiveness}. However, work that examines how individual differences affect interactions with and perceptions of LLM-powered tools in educational settings is sparse, even though understanding the role of demographic and individual variability is crucial~\cite{zhang2024profiling}. This is particularly important for developing inclusive and effective educational tools that suit the diverse needs of students.

Our work seeks to address these research gaps by conducting a field study that evaluates the use of LLM-powered tools for an extended period of time. Particularly, our study not only aims to evaluate the practicality of LLMs in programming learning educational contexts, but also intends to contribute to a more nuanced understanding of their long-term implications for learning and teaching methodologies. 

\section{Method} 
In this section, we describe the design of CodeTutor (\autoref{subsec:method_design_codetutor}), an overview of our participants (\autoref{subsec:method_participant}), our study procedure and data collection (\autoref{subsec:method_study_flow}), and our quantitative and qualitative data analysis (\autoref{subsec:method_data_analysis}). The source code of CodeTutor, pre-test questions, and data analysis code is available on \href{https://osf.io/e3zgh}{osf.io/e3zgh}.

\subsection{Design of CodeTutor}
\label{subsec:method_design_codetutor}

\begin{figure*}[h]
\centering
\includegraphics[width=.85\linewidth]{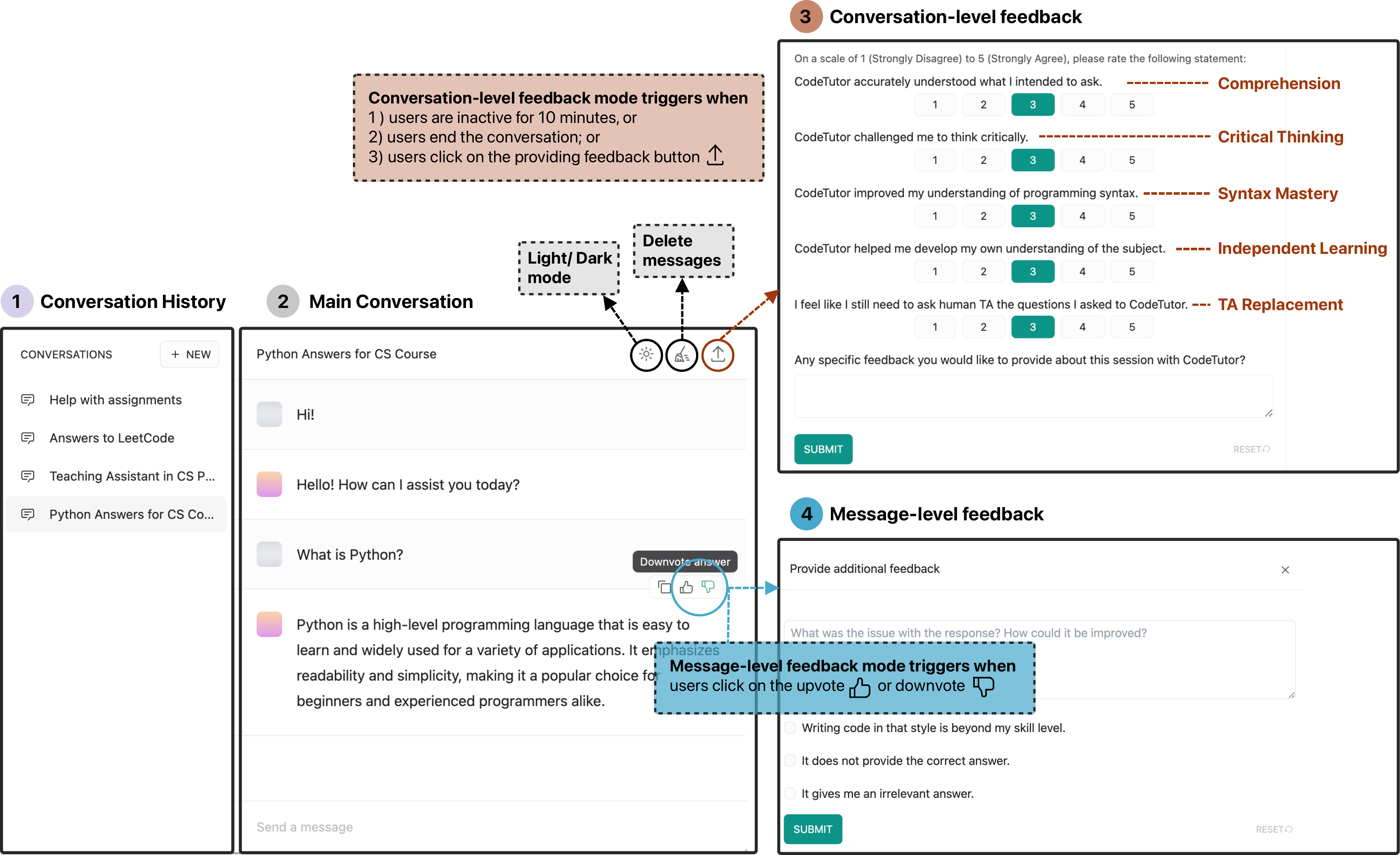}
\caption{CodeTutor is a web application that leverages OpenAI API, featuring four main components: \includegraphics[height=3mm]{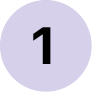} Conversation History that lists different conversation threads, \includegraphics[height=3mm]{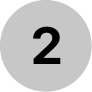} Main Conversation that shows an ongoing dialogue with CodeTutor, \includegraphics[height=3mm]{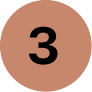} Conversation-level Feedback module that allows users to elaborate on their attitudes towards CodeTutor by proving ratings on 1) comprehension, 2) critical thinking, 3) syntax mastery, 4) independent learning, and 5) TA replacement likelihood, and to provide specific comments, and \includegraphics[height=3mm]{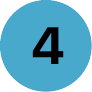} Message-level Feedback that offers options for users to give detailed feedback on individual messages or responses from CodeTutor.}
\label{fig:CodeTutor_UI}
\end{figure*}

We developed CodeTutor, a browser-based web application using TypeScript and front-end frameworks (e.g., SolidJS, Astro, and libraries such as Zag), for a responsive and interactive user interface. CodeTutor integrates OpenAPI API, which enables the GPT-3.5 model offered by OpenAI. The main interface is shown in \autoref{fig:CodeTutor_UI}. \\
\textbf{Login. }  
Students log in to CodeTutor using their email addresses, with a randomly generated unique identifier (UID) that tracks their activities anonymously. \\
\textbf{User Interface. }  
The CodeTutor interface features a navigation sidebar and a central chat area. The sidebar enables easy navigation, with a button for starting new conversations and a chronological listing of existing ones for quick access. \\
\textbf{User Feedback Structure. }  
Feedback is important in CodeTutor in order to understand user engagement and students' attitudes towards it. CodeTutor provides two feedback mechanisms: 1) conversation-level and 2) message-level feedback. \\
\textbf{Data Storage. }
CodeTutor stores data locally on the user's browser with IndexedDB and can only \textit{upload} essential information with our secure server for research purposes, where a unique ID for anonymous tracking identifies each conversation. To protect privacy, CodeTutor cannot read stored data from our server. \\
\textbf{API Usage.}
OpenAI only offered limited configuration ability for their API at the time we started our experiment. So we carefully crafted the \textit{system role} text in our implementation to specify the model to answer questions as a teaching assistant in an entry-level Python class, making answers from OpenAI API consistent even if the length of a conversation exceeds its token limit.

\subsection{Participants}
\label{subsec:method_participant}
Upon approval from our institution's Institutional Review Board (IRB), we conducted a field study evaluation study with 50 participants. The field study took place in the Computer Science Department of a 4-year university in the United States. Our criteria for participation include: Participants need to be 18 years or older, be able to speak and write in English, and register as entry-level undergraduate computer science students at our institution. \autoref{tab:demographic} presents an overview of our participants' demographic information.

 \begin{table}[h]
  \caption{Overview of Participant Characteristics} 
  \label{tab:demographic} 
  \footnotesize
  \begin{tabular}{llr}
    \toprule
    \textbf{Characteristics}  & \textbf{Options} & \textbf{Number of } \\ 
    & & \textbf{participants} \\
    \midrule
    Gender      & Woman         & 22   \\
                & Man           & 25  \\
                & Non-binary    & 1   \\
                & Prefer not to say & 2  \\
    \midrule  
    Major   & Computer Science  & 18   \\
            & Data Science      & 9 \\
            & Biology           & 5   \\
            & Mathematics       & 4   \\
            & Economics         & 3  \\
            & Others            & 10  \\
            & Not reported      & 1 \\
    \midrule  
    Year of Study   & Freshman  & 37  \\
                    & Sophomore & 5  \\
                    & Junior    & 6  \\
                    & Senior    & 1  \\
                    & Not reported & 1  \\     
    \midrule
    Race    & African American or Black & 1   \\
            & Asian        & 17 \\
            & Multiracial  & 3   \\
            & White        & 26   \\
            & Not reported & 3  \\
    Ethnicity    & Latino/Hispanic & 3   \\
    \midrule
    Prior Experience  & Only ChatGPT & 28   \\
    with LLM tools   & ChatGPT and other tools & 11   \\
                     & Never used & 11  \\
    \bottomrule
  \end{tabular}
\end{table}

\subsection{Study Procedure \& Data Collection}
\label{subsec:method_study_flow}
Our field study lasted from September 27 (after the course add-drop period) to December 11, 2023 (the final exam due). Below, we describe each component of our study. 

\subsubsection{Pre-test} 
Participants were initially requested to provide their consent to participate, with being informed about the study's objectives, procedures, and their rights as participants, including the right to withdraw at any time without penalty. Following the consent process, the pre-test assessment was administered to evaluate students' existing knowledge of Python programming, providing a baseline for subsequent analysis.

This pre-test included three sections with Python questions, with a total of 22 questions that varied in difficulty for an evaluation of participant skills. The first section featured eight questions (Questions 1-8, for example, \textit{``What is the output of the following code: print(3+4)?''} ), the second section included seven questions of medium difficulty (Questions 9-15, for example, \textit{``If I wanted a function to return the product of two numbers a and b, what should the return statement look like?''}), and the third section presented seven challenging questions (Questions 16-22, for example, \textit{``What will be the output of the following code? [Multiple lines of code]''}). The total score of the three sections was 100 points. Pre-test submissions were graded by our researchers with Computer Science backgrounds, using predetermined scoring criteria.

This pre-test also asked about participants' prior experience with LLMs, specifically asking, \textit{``Which of the following Large Language Model AI tools have you used before? Please select all that apply.''} Participants were also asked to provide demographic information, including their major (or intended major), gender, and race/ethnicity. 
Participants were assured that all demographic information would remain anonymous and be used solely for research purposes. 

\subsubsection{Control vs. Experimental Group}
Participants were divided into two groups: the control group, which used traditional learning methods and had access to human teaching assistants (TAs) for additional support outside class hours, and the experimental group, which used CodeTutor as their primary educational tool beyond class hours, alongside access to standard learning materials and human TAs.  
Using LLM-based tools other than CodeTutor in this course was prohibited.  

\begin{figure}[h!]
\centering
\includegraphics[width=.85\linewidth]{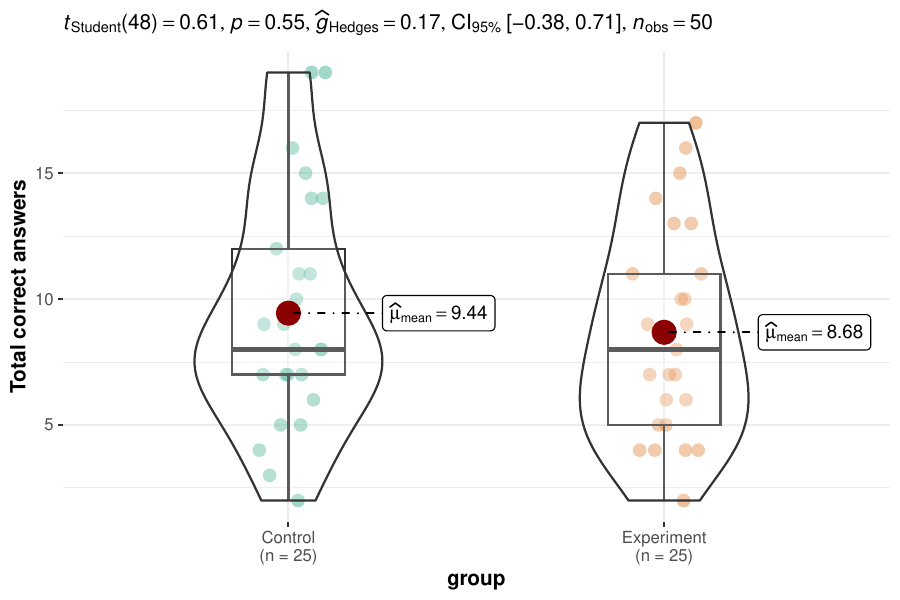}
\caption{Parametric pairwise comparison (ANOVA) reveals no significant difference in correct answer count of pre-test in the control and experimental groups.} 
\label{fig:pretest_score}
\end{figure}

To divide participants into a control group and an experimental group, we initially sorted the entire sample based on their previous engagement with LLM-powered tools, resulting in two groups: those who have used any LLM-powered tools before (\textit{Used Before}) and those who have not (\textit{Never Used}). Within the \textit{Used Before} category, we split the participants into two subsets, \textit{Used Before Subset A} and \textit{Used Before Subset B}, based on the overall pre-test result distribution to ensure both subsets are representative of the wider group. The same process was applied to the \textit{Never Used} group, generating two additional subsets: \textit{Never Used Subset A} and \textit{Never Used Subset B}. The experimental group is then formed by combining \textit{Used Before Subset A} with \textit{Never Used Subset A}, while the control group consists of the combination of \textit{Used Before Subset B} and \textit{Never Used Subset B}. This method ensures the experimental and control groups were balanced regarding prior experience with Chatbots and their pre-test performance (see \autoref{fig:pretest_score}).

Following their group assignments, students in the experimental group were sent detailed instructions via email on how to access and use CodeTutor. In the field study, participants were not mandated to adhere to a specific frequency of engagement with CodeTutor; instead, they were encouraged to utilize the tool at their own pace. This approach allowed for a naturalistic observation of how students integrate LLM-powered educational resources into their learning processes, without imposing additional constraints that could influence their study habits or the study's outcomes.

\subsubsection{Student Evaluation} 
At the end of the semester, students' final grades were used as a primary measure to assess their learning outcomes and the impact of CodeTutor interventions. While acknowledging that final grades are influenced by various factors, they offer a standardized measure of overall academic success, enabling an assessment of CodeTutor's role in improving student learning outcomes. 

Final grades were determined by a weighted average that includes several components for each student: labs (practical mini-projects), assignments (individual coding tasks, such as array summation), mid-terms, and a final exam (comprising questions similar to those in the pre-test). Note that a student's final grade can surpass 100 if bonus points are awarded throughout the semester. Access to CodeTutor is restricted during mid-terms and final exams, categorizing the assessment components into two groups: \textit{CodeTutor-Allowed} (labs and assignments) and \textit{CodeTutor-Not-Allowed} (mid-terms and final exams). This categorization facilitates an analysis of CodeTutor's impact on student performance by examining potential dependencies on the tool and the improvement of learning outcomes in its absence. 

\subsection{Data Analysis}
\label{subsec:method_data_analysis}
\subsubsection{Quantitative Data Analysis}
We examined the students' scores, interaction behaviors, and attitudes of using CodeTutor through multiple statistical analyses. 

First, we calculated descriptive statistics for all variables, including frequency with percentage for categorical variables and means and standard deviations for continuous variables. To examine the variation in students' scores before and after the intervention (i.e., the use of CodeTutor), we conducted \emph{paired-t tests} for both the experimental and control groups. \emph{Multiple regression analyses with family-wise p-value adjustment} were used to examine the effects of CodeTutor on score improvement, taking into account students' past experiences using LLM-powered tools and demographic variables, such as major, gender, and race. We then investigated the impact of CodeTutor accessibility on academic performance with \emph{ANOVA method}. Moreover, we conducted a \emph{chi-squared test} to explore the relationship between the quality of students' content and prompts and CodeTutor performance. To understand students' attitudes towards CodeTutor, we calculated \emph{Spearman's correlation matrix for continuous variables}, given the characteristics of our data, which are non-normal and exhibit unequal variance. Furthermore, to examine differences between questions, we used the \emph{Kruskal-Wallis Rank Sum Test} (using R package stats~\cite{statscite}) and then performed post-hoc tests using \emph{Dunnett's test} (using the R package FSA~\cite{fsacite}) in cases where significant differences were found. To investigate the importance of time on students' attitudes towards CodeTutor, we introduced a \emph{linear mixed effects} (LME) model (using the R package lme4~\cite{lme4cite}). We considered statistical significance at a significance level of $p <$ 0.05 for most cases, except in multiple regression analyses where we used $p <$ 0.1 and showed effect sizes were significant enough to indicate the relationship of variables.

\subsubsection{Qualitative Data Analysis}
We also analyzed the conversational history between users. Specifically, we used the \emph{General Inductive Approach}~\cite{thomas2006general} to guide our thematic analysis of the conversational data. The first author conducted a close reading of the data to gain a preliminary understanding of the conversational data and then labeled the text segments to formulate categories, which served as the basis for constructing low-level codes to capture specific elements of the user-CodeTutor interactions. Similar low-level codes were then clustered together to achieve high-level themes. During the analysis, the research team engaged in ongoing discussions to refine and clarify emerging themes.

\section{Results}   

In this section, we examined the impact of CodeTutor on student academic performance (\autoref{sec:score_improvements} to answer RQ1), analyzed students' attitudes towards learning with CodeTutor (\autoref{sec:students_attitudes} to answer RQ2), and characterized their engagement patterns in entry-level programming courses (\autoref{sec:students_engagement} to answer RQ3). 

\subsection{RQ1: Learning Outcomes with CodeTutor}
\label{sec:score_improvements}

\subsubsection{Comparative Analysis of Score Improvements}
Overall, students in the experimental group exhibited a greater average improvement in scores, as illustrated by comparing their pre-test and final scores to those in the control group. Specifically, the average increase for the experimental group was 12.50, whereas the control group showed an average decrease of 3.17 when comparing final scores to pre-test scores.

We conducted paired t-tests for both the experimental and control groups to determine if the observed improvements were statistically significant, starting with the premise that there were no differences in pre-test scores between these two groups. Our null hypothesis assumed that the true mean difference between pre-test and final scores was zero. For the control group, the null hypothesis could not be rejected, suggesting that the differences between pre-test and final scores were not statistically significant ($t =$ -0.879, $p =$ 0.394). Conversely, participants in the experimental group demonstrated significant improvement from the pre-test to final scores, indicating a statistically significant enhancement in their scores ($t =$ -2.847, $p =$ 0.009). 

Furthermore, when examining the improvement in \textit{CodeTutor-Not-Allowed} components, the experimental group exhibited an average increase of 7.33, whereas the control group showed no significant change. By conducting a paired t-test comparing the pre-test and final exam scores (during which the use of CodeTutor was not permitted), it was observed that students in the experimental group demonstrated a statistically significant improvement ($t =$ -2.405, $p =$ 0.026). This result suggests that students who have used CodeTutor exhibit more substantial improvement even when CodeTutor is unavailable.

\subsubsection{Effect of CodeTutor Accessibility on Academic Performance}

\begin{figure}[h]
\centering
\includegraphics[width=.85\linewidth]{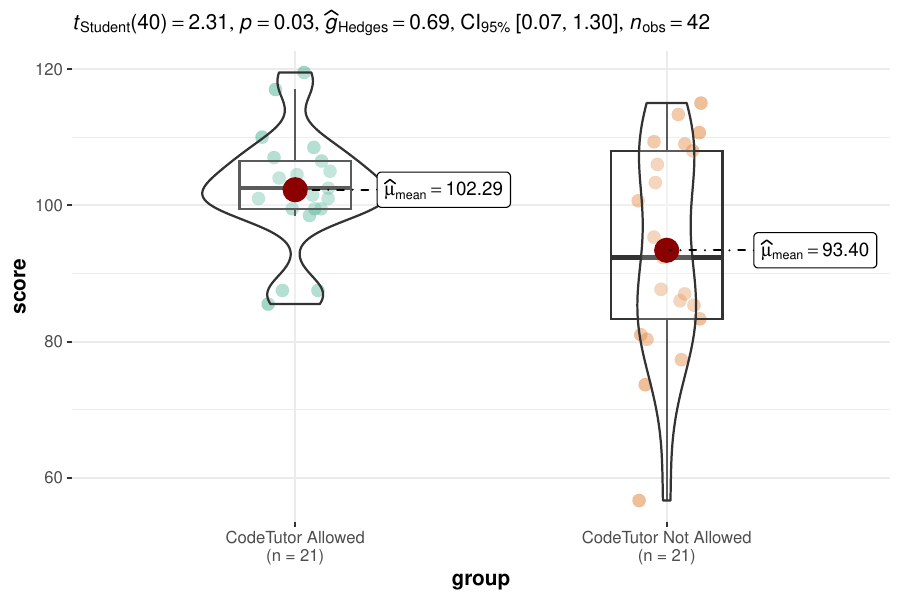}
\caption{Parametric pairwise comparison (ANOVA) reveals a significantly higher mean score in the ``CodeTutor-Allowed'' group compared to the ``CodeTutor-Not-Allowed'' group.}
\label{fig:anova}
\end{figure}

By constructing the \textit{CodeTutor-Allowed} and \textit{CodeTutor-Not-Allowed}, we determine the correlation between CodeTutor's accessibility and student academic performance. Using the ANOVA technique on the data from the experimental group, \autoref{fig:anova} reveals that the mean score for the \textit{CodeTutor-Allowed} category stands at 102.29, in contrast to the \textit{CodeTutor-Not-Allowed} components, which has a mean score of 93.40. The statistical analysis results show a significant difference between the two groups ($t =$ 2.31, $p =$ 0.03), suggesting that the allowance of CodeTutor correlates with higher student scores.

\subsubsection{Correlation Between Student Demographics and Final Scores in the Experimental Group} 

\begin{table}
\centering
\caption{Multiple regression models explaining respondents' \textbf{final scores} in experimental group. (Significance level: $^\dag$ $p < 0.1$, * $p < 0.05$, ** $p < 0.01$, *** $p < 0.001$). }
\label{tab:demographic_multiple_regression}
\resizebox{0.47\textwidth}{!}{
\begin{tabular}{lllll}
\toprule
& Estimate & Std. Error & t value & Pr($>$$|$t$|$) \\ 
\midrule
Const & 93.683 & 3.877 & 24.166 & 0.000 *** \\ 
\cdashlinelr{1-5}
\textbf{Prior Experiences with LLM tools} \\
(\textit{Reference: Used before}) \\ 
Never used & 18.877 & 5.054 & 3.735 & 0.032 * \\ 
\cdashlinelr{1-5}
\textbf{Major} \\
(\textit{Reference: Computer science}) \\ 
Data Science & 14.532 & 5.662 & 2.567 & 0.073 $^\dag$ \\ 
Mathematics & 17.692 & 5.852 & 3.023 & 0.057 $^\dag$ \\ 
Biology & 16.257 & 5.662 & 2.871 & 0.057 $^\dag$ \\ 
Economics & 1.362 & 4.799 & 0.284 & 0.784 \\ 
Others & -13.004 & 6.022 & -2.160 & 0.115 \\ 
\cdashlinelr{1-5}
\textbf{Gender} \\
(\textit{Reference: Female}) \\ 
Male & 5.917 & 3.845 & 1.539 & 0.223 \\
\cdashlinelr{1-5}
\textbf{Race} \\
(\textit{Reference: White}) \\ 
Asian & -7.831 & 3.933 & -1.991 & 0.128 \\ 
African American or Black & 8.099 & 7.107 & 1.140 & 0.322 \\ 
Others & 6.102 & 5.416 & 1.127 & 0.322 \\ 
  \bottomrule
\end{tabular}
}
\end{table}

Subsequently, we evaluated demographic factors to determine whether specific student groups, particularly those with prior tech experience, experienced greater benefits from CodeTutor.
\autoref{tab:demographic_multiple_regression} shows the results of multiple regression models, examining how students' final scores in the experimental group are associated with their LLM history, major, gender, and race. Students who have never used any LLM-powered tools performed a significant increase ($\beta =$ 18.877, $p =$ 0.032) in final score than the students who used it before. 

Moreover, differences in final scores among various majors within the experimental group were statistically significant, indicating that majors play a substantial role in final scores in the experimental group. Students majoring in data science ($\beta =$ 14.532, $p =$ 0.073), mathematics ($\beta =$ 17.692, $p =$ 0.057), and biology ($\beta =$ 16.257, $p =$ 0.057) exhibited a significant positive correlation with final scores compared to those majoring in computer science, suggesting that these majors achieved higher final scores. In terms of gender, no significant effects were observed, indicating no difference between genders in final scores. Additionally, no significant differences were noted across the races in final scores.

\textbf{\underline{Summary of results of RQ1: }}
Collectively, our findings suggest that students in the experimental group achieved significant score improvements with CodeTutor. Particularly, those who were new to CodeTutor achieved even greater improvements, while students majoring in data science, mathematics, and biology surpassed their computer science counterparts. Moreover, students exhibited higher scores when permitted to use CodeTutor. 

\subsection{RQ2: Students' Attitudes towards CodeTutor}
\label{sec:students_attitudes}

\subsubsection{Descriptive Analysis}

\begin{figure}[h]
\centering
\includegraphics[width=\linewidth]{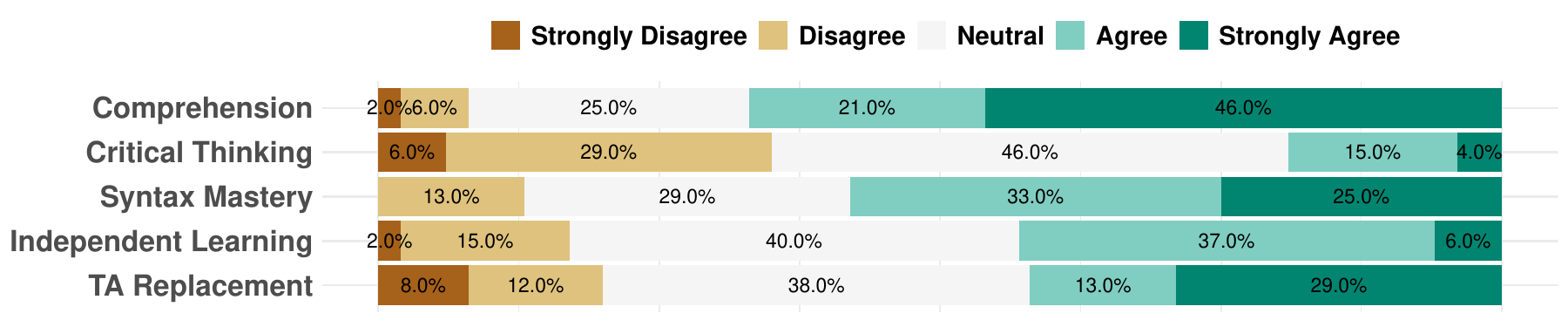}
\caption{Participants’ attitudes toward CodeTutor, in terms of comprehension, critical thinking, syntax mastery, independent learning, and TA replacement (see \autoref{fig:CodeTutor_UI} for detailed questions). }
\label{fig:barplot}
\end{figure}

In terms of students' attitudes towards CodeTutor (see \autoref{fig:CodeTutor_UI} \includegraphics[height=2.5mm]{figures/icon3.pdf} for the specific questions), we found that a small portion of students (8\%) strongly disagreed or disagreed that \emph{CodeTutor accurately understood what students intended to ask}, while most (67\%) agreed or strongly agreed. In addition, 35\% strongly disagreed or disagreed that \emph{CodeTutor helped them think critically}, while 19\% agreed or strongly agreed. Furthermore, 13\% students disagreed that \emph{CodeTutor improved their understanding of programming syntax}, with a larger proportion of individuals agreeing (33\%) or strongly agreeing (25\%). Nearly half of the students (42\%) agreed or strongly agreed that \emph{CodeTutor helped students build their own understandings}, while very few (17\%) strongly disagreed or disagreed. Finally, regarding \emph{the potential of CodeTutor to substitute for a human teaching assistant}\footnote{In our analysis, response values to the TA Replacement question were reversed, so a higher score indicates a stronger preference for our tool over human teaching assistants. This reversal is consistently applied across all subsequent analyses.}, 20\% of the students strongly disagreed or disagreed with this notion, while 42\% of them agreed or strongly agreed. \autoref{fig:barplot} shows the distribution of students' responses across these five questions.

\subsubsection{Exploring Relationships in Student Attitudes Toward CodeTutor}

\begin{figure}[h]
\centering
\includegraphics[width=0.75\linewidth]{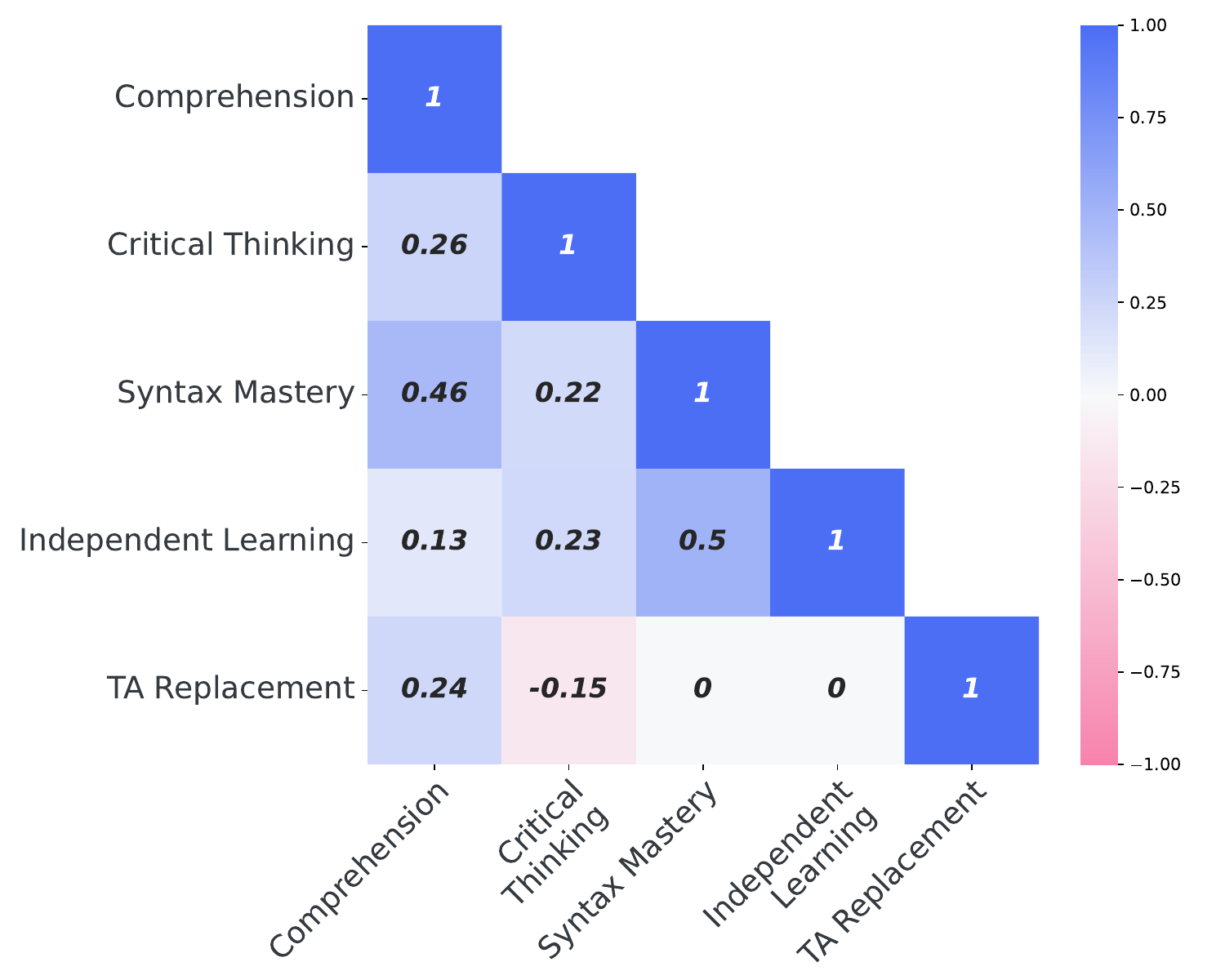}
\caption{A correlation matrix heatmap visualizing the relationship between different metrics. The blue color indicates positive correlations, while pink represents negative correlations. Correlation coefficients are displayed inside each cell.} 
\label{fig:corr}
\end{figure}

\begin{figure}[h]
\centering
\includegraphics[width=.85\linewidth]{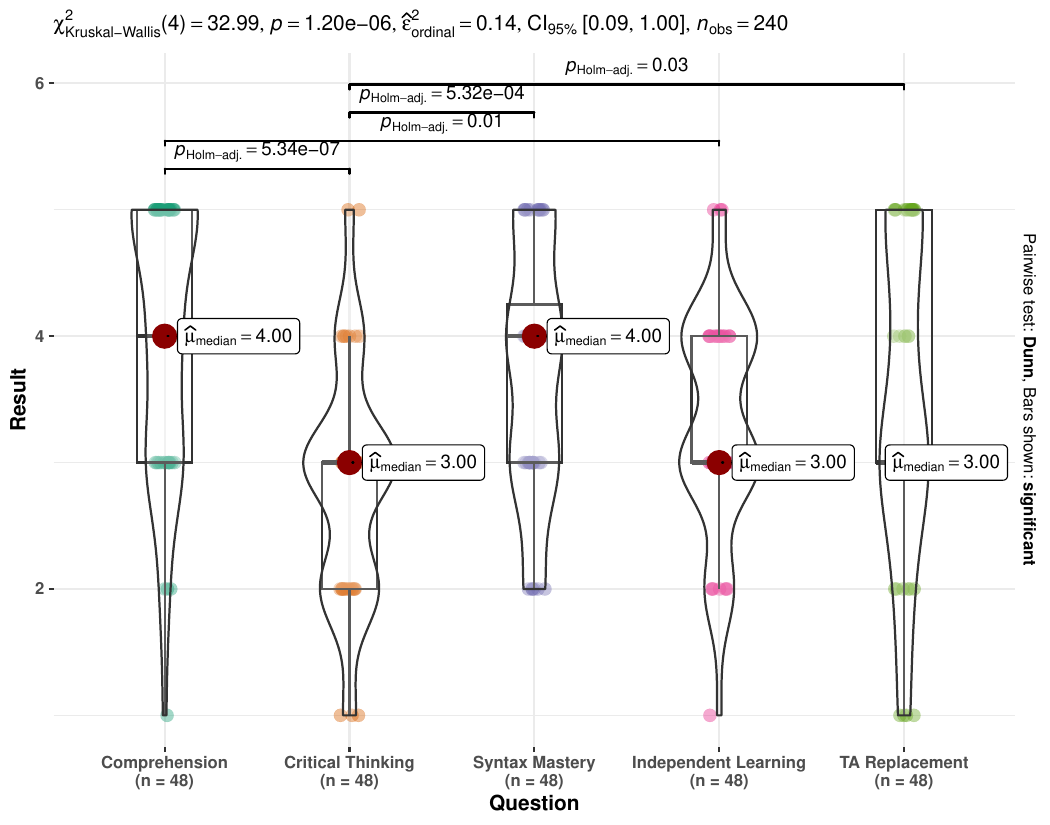}
\caption{Non-parametric pairwise comparison test (Dunn's test): Differences in agreement levels across different questions. We can see that students predominantly favored CodeTutor for its comprehension and syntax support rather than its ability to foster critical thinking. Additionally, there was a stronger consensus on CodeTutor's proficiency in understanding queries compared to its effectiveness in enhancing programming syntax.} 
\label{fig:dunntest}
\end{figure}

\begin{table*}
\footnotesize
\centering
\caption{Linear Mixed-Effects Model of Student Attitudes over time. (Significance level: $^\dag$ $p < 0.1$, * $p < 0.05$, ** $p < 0.01$, *** $p < 0.001$). Over time, students exhibit a significant decline in their agreement with CodeTutor's comprehension and replacement of human teaching assistants.}
\label{tab:lme}
\begin{tabular}{llllll}
\toprule
& Comprehension & Critical Thinking & Syntax Mastery&Independent Learning &TA Replacement\\ 
&$\beta$ (\textit{Std. Error}) &$\beta$ (\textit{Std. Error})&$\beta$ (\textit{Std. Error}) &$\beta$ (\textit{Std. Error}) &$\beta$ (\textit{Std. Error}) \\
\midrule
Const & 4.700(0.297)*** & 2.690(0.247)*** &3.760(0.262)***&3.044(0.218)***&3.964(0.330)***\\ 
Time & -0.114(0.039)** & 0.040(0.037)&-0.018(0.041)&0.054(0.036)&-0.099(0.051)$^\dag$\\
  \bottomrule
\end{tabular}
\end{table*}

\autoref{fig:corr} reveals key relationships among students' attitudes on CodeTutor. The moderate positive correlation between Comprehension and Syntax Mastery suggests that proficiency in one is associated with higher performance in the other. Critical Thinking is slightly positive with Comprehension and Independent Learning but slightly negative with TA Replacement. Furthermore, Syntax Mastery strongly correlates with Independent Learning, indicating a close relationship between mastering programming syntax and self-directed learning outcomes. In addition, TA Replacement has minimal to no significant correlations with other variables, suggesting its effects vary independently of these educational aspects.

To further explore the relationship of different students' attitudes among questions, we present the results of multiple comparisons across the five questions. Specifically, our results show that respondents' attitudes ($\chi^2 =$ 32.99, $p <$ 0.05) significantly differ across questions. Our post-hoc tests (see \autoref{fig:dunntest}) further reveal that students were significantly less in agreement about CodeTutor's assistance in fostering critical thinking compared to its ability to understand, help in learning syntax and serving as a replacement for a teaching assistant. Moreover, our findings suggest that respondents were significantly more in agreement with CodeTutor's effectiveness in comprehension than in its ability to improve students' understanding of programming syntax.

We then conducted a linear mixed effects (LME) model to explore time's influence on students' attitudes toward CodeTutor:
$$
QuestionIndicator_{it}=\beta_0+b_{0i}+(\beta_1+b_{1i})t+\epsilon_{it}
$$
where $\beta_0$ and $\beta_1$ are unknown fixed effect parameters; $b_{0i}$ and $b_{1i}$ are the unknown student-specific random intercept and slope, respectively, which are assumed to have a bivariate normal distribution with mean zero and covariance matrix $D$; $QuestionIndicator$ is the student response at time $t$; and $\epsilon_{it}$ is the residual error for student $i$ at time $t$, with a normal distribution $N(0,\sigma^2)$, which is assumed to be independent of the random effects. From \autoref{tab:lme}, we can see that students' attitudes toward CodeTutor show a significant decrease in Comprehension ($\beta =$ -0.114, $p <$ 0.01), which indicates that students disagree with CodeTutor's understanding accuracy over time. Moreover, there is a weakly significant decrease in TA Replacement ($\beta =$ -0.099, $p <$ 0.1) with increasing time. This shows a slight tendency for them to consider more human TA help over time. Also, students perform no significant difference over time in Critical Thinking, Syntax Mastery, and Independent Learning.

\textbf{\underline{Summary of results of RQ2: }}
In summary, students recognize CodeTutor's ability to understand their queries and assist with programming syntax yet question its capacity to promote critical thinking skills. Additionally, students' confidence in CodeTutor's comprehension abilities decreases over time, with a growing preference for support from human teaching assistants.

\subsection{RQ3: Students' Engagement with CodeTutor}
\label{sec:students_engagement}

In total, we documented 82 conversation sessions\footnote{In our analysis, a conversation session is a continuous exchange of messages between users and CodeTutor within a specific period, characterized by a coherent topic or purpose.} with CodeTutor, encompassing a total of 2,567 messages. In these sessions, 415 unique topics were discussed, averaging 5.06 topics per session and 6.19 messages per topic.

\subsubsection{Message Classification \& Interaction Patterns}
In total, we collected 2567 conversational messages exchanged between users and the CodeTutor. Of these, 1288 messages originated from the users, and CodeTutor responded with 1279 messages.  

\autoref{tab:definition_categories} presents categorizations of messages between users and CodeTutor. Each category has a description and an example to illustrate the message type. Categories of messages from both users \includegraphics[height=2.5mm]{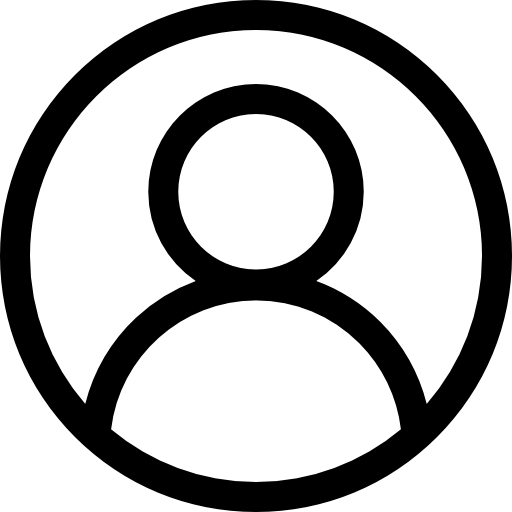} and CodeTutor \includegraphics[height=2.5mm]{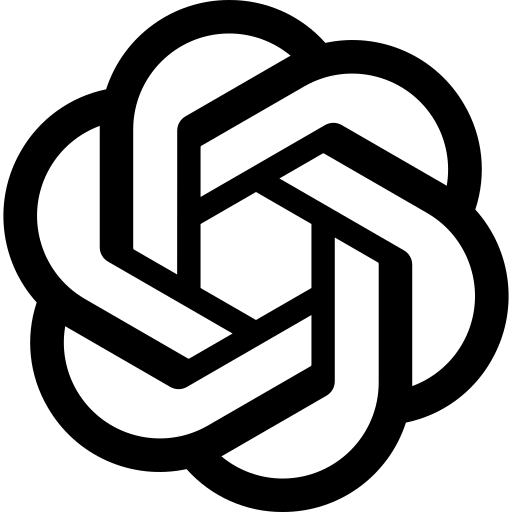} include \textit{Programming Task inquiries}, addressing specific Python programming challenges; \textit{Grammar and Syntax} questions, focusing on Python's basic grammar or syntax without necessitating runnable programs; \textit{General Questions}, which are not directly related to Python; and \textit{Greetings}, initiating or finishing interaction.

From the users' side \includegraphics[height=2.5mm]{figures/user.png}, additional categories highlight their engagement with CodeTutor: \textit{Modification Requests} for alterations to previous answers; \textit{Help Ineffective} indicating issues or errors in CodeTutor's provided solutions; \textit{Further Information} to elaborate on prior queries; and \textit{Debug Requests} for assistance in resolving bugs or errors in code snippets.

CodeTutor's responses \includegraphics[height=2.5mm]{figures/gpt.png} are classified into \textit{Corrections}, which address and amend errors in previous responses and \textit{Explanations}, providing further details on provided solutions or clarify why certain requests cannot be fulfilled.

 \begin{table*}[h]
  \caption{Categorizations of messages, from users' side \includegraphics[height=3mm]{figures/user.png} and from CodeTutor's side \includegraphics[height=3mm]{figures/gpt.png}. \textit{[Code Snippet]} represents a Python code segment. The \textit{Percentage} column represents the ratio of occurrences of each category to the total number of messages. Note that some categories may only apply to messages sent by either users or CodeTutor, and messages may carry multiple categories.} 
  \label{tab:definition_categories} 
  \footnotesize
  \begin{tabular}{lp{5cm}p{5cm}r}
    \toprule
    \textbf{Category Name}  & \textbf{Description} 
    & \textbf{Example}      & \textbf{Percentage}\\ 
    \midrule
    \raisebox{-.25\height}{\includegraphics[height=3mm]{figures/user.png}} 
    \raisebox{-.25\height}{\includegraphics[height=3mm]{figures/gpt.png}}
    Programming Task  
    & Any questions or answers related to Python programming. 
    & \textit{``Write a function that prints the nth(argument) prime number.''}  
    & 86.52\%
    \\ 
    \raisebox{-.25\height}{\includegraphics[height=3mm]{figures/user.png}} 
    \raisebox{-.25\height}{\includegraphics[height=3mm]{figures/gpt.png}}
    Grammar \& Syntax
    & When a message is related to basic Python grammar or syntax problems, a runnable program is most likely unnecessary.
    & \textit{``What does \{\} do in Python?''}
    & 14.26\%
    \\ 
    \raisebox{-.25\height}{\includegraphics[height=3mm]{figures/user.png}} 
    \raisebox{-.25\height}{\includegraphics[height=3mm]{figures/gpt.png}} 
    General Question
    & When a message is not directly related to Python.
    & \textit{``What is ASCII?''}
    & 4.29\%
    \\ 
    \raisebox{-.25\height}{\includegraphics[height=3mm]{figures/user.png}} 
    \raisebox{-.25\height}{\includegraphics[height=3mm]{figures/gpt.png}}
    Greetings
     & When a message is greeting.
    & \textit{``Hello! How can I assist you today?''}
    & 0.62\%
    \\
    \midrule  
    \raisebox{-.25\height}{\includegraphics[height=3mm]{figures/user.png}} 
    Help Ineffective
    & When a \textbf{user} message says the previous answer generated by CodeTutor is wrong or provides error information.
    & \textit{``This code still fails.''}  
    & 12.86\%
    \\ 
    \raisebox{-.25\height}{\includegraphics[height=3mm]{figures/user.png}}
    Debug Request
    & When a \textbf{user} message asks CodeTutor to fix bugs or explain what was wrong in code snippets provided or in previous messages.
    & \textit{``Debug this code. [Code Snippet]''}  
    & 8.22\%
    \\ 
    \raisebox{-.25\height}{\includegraphics[height=3mm]{figures/user.png}}
    Modification Request
    & When a \textbf{user} requires CodeTutor to change something on its previous answer.
    & \textit{``Remove comments.''}
    & 4.48\%
    \\  
    \raisebox{-.25\height}{\includegraphics[height=3mm]{figures/user.png}}
    Further Information
    & When a \textbf{user} message provides more context on their previous input.
    & \textit{``All the input strings will be the same length.''}  
    & 3.97\%
    \\ 
    \midrule   
    \raisebox{-.25\height}{\includegraphics[height=3mm]{figures/gpt.png}}
    Explanation
    & When \textbf{CodeTutor} explains something in previous messages or why it cannot complete the current task from users.
    & \textit{``I'm sorry, but I need more information to provide the answers for questions 4 and 6.''}
    & 28.94\%
    \\
    \raisebox{-.25\height}{\includegraphics[height=3mm]{figures/gpt.png}}
    Correction
    & When \textbf{CodeTutor} corrects content in its previous answer. 
    & \textit{``Apologies for the syntax error. Here is the corrected version: [Code Snippet]''} 
    & 13.95\%
    \\ 
    \bottomrule
  \end{tabular}
\end{table*}

\subsubsection{Analysis of Prompt Quality \& Correlation with Response Effectiveness}
\label{sec:prompts_quality}
To further examine user interaction patterns with CodeTutor and their implications for its educational value, we analyzed the relationship between prompt quality and response accuracy. This analysis stems from the premise that detailed and precise prompts are likely to improve the AI's understanding of user requirements, thereby potentially raising the standard of its responses.

To do so, we evaluated a corpus of 1,190 prompts, after removing all greeting messages, to assess their quality. Our analysis showed that 37\% were deemed good quality. The remaining 63\% were identified as poor quality. We defined ``good quality'' prompts as providing sufficient detail for CodeTutor to generate an accurate response. In contrast, ``poor quality'' prompts were those that did not meet this criterion. We categorized the deficiencies in poor quality prompts into four types: incomplete information ($n =$ 189, 25\%), which lacked specific details necessary for CodeTutor to understand the context; lack of clear goals ($n =$ 172, 23\%), where the desired outcome was not explicitly stated; over-reliance on CodeTutor ($n =$ 362, 48\%), where the assignment questions are directly copied and pasted into CodeTutor; and poor structural organization ($n =$ 25, 3\%), which exhibited unclear or confusing request structures. Prompts were further labeled as ``working'' if they elicited an appropriate response from CodeTutor, and ``not working'' if they failed to do so.

Using a Chi-square test, we investigated whether the prompt quality and the effectiveness of CodeTutor's responses were independent. Our results showed a significant correlation ($\chi^2 =$ 144.84, $p <$ 0.001). In other words,  clearer and more detailed prompts are associated with responses that are more likely to be effective. 

\textbf{\underline{Summary of results of RQ3: }} 
We characterized the messages exchanged between users and CodeTutor. We categorize these interactions between users and CodeTutor into inquiries (e.g., programming tasks, syntax questions) and feedback alongside CodeTutor's responses (corrections and explanations), illustrating a dynamic exchange aimed at facilitating learning. We also found that the clarity and completeness of prompts are significantly correlated with the quality of responses from CodeTutor.

\section{Discussion} 

Our semester-long field study provided insights into how students in introductory computer science courses utilized CodeTutor and its effects on educational outcomes. In short, our results show that 1) students who used CodeTutor had shown significant improvements in scores; 2) while CodeTutor was valued for its assistance in comprehension and syntax, students expressed concerns about its capacity to enhance critical thinking skills; 3) skepticism regarding CodeTutor as an alternative to human teaching assistants grew over time; 4) CodeTutor was primarily used for various coding tasks, including syntax comprehension, debugging, and clarifying fundamental concepts; 5) the effectiveness of CodeTutor responses was notably higher when prompts were clearer and more detailed. Building on these findings, we discuss the implications for future enhancements and research directions in the rest of the section.

\subsection{Towards Enhancing Generative AI Literacy} 
Our research indicates a positive correlation between the use of Generative AI tools and improved student learning outcomes. However, 63\% of student-generated prompts were deemed unsatisfactory, indicating a lack of essential skills to fully exploit Generative AI tools. This finding also suggests the need to promote \textit{Generative AI literacy} among students. Here, we define \textit{Generative AI literacy} as the ability to effectively interact with AI tools and understand how to formulate queries and interpret responses. Our findings suggest that while students can leverage CodeTutor for practical coding assistance and syntax understanding, there is a gap in using these tools to enhance critical thinking skills. We suggest educational programs integrate Generative AI literacy as a core component of their curriculum, teaching students how to use these tools for immediate problem-solving and engaging with them to promote deeper analytical and critical thinking. This could include workshops on effective query formulation, sessions on interpreting AI responses, and exercises designed to challenge students to critically evaluate the information and solutions offered by AI tools.  

We also propose approaches to integrate HCI tools and principles into LLM-enabled platforms, such as \textit{prompt construction templates} providing users with templates or structured forms for crafting queries. They can guide users in formulating more effective and precise questions. Templates could include placeholders for essential details and context, providing the necessary information for the AI to generate accurate responses to users. Furthermore, integrating \textit{Critical Thinking Prompts} might be particularly effective in stimulating in-depth analytical thinking. For example, the interface could pose follow-up questions encouraging users to assess AI answers' adequacy critically. Questions such as, ``Does this response fully address your query?'' or ``What additional information might you need?'' may prompt users to engage in a more thorough evaluation of the information provided, fostering a habit of critical reflection and assessment. Another possible approach is \textit{Facilitating Collaborative Query Building}, which leverages the power of collective intelligence. By designing interfaces that support real-time collaboration among users, individuals can work together to construct and refine queries. We can also use LLMs to evaluate and refine user questions instantly as they perform well in prompting~\cite{zhou2022large}.
 
\subsection{Turning to the Temporal Dynamics of LLM-Powered Tutoring Tools} 
The temporality aspect of using CodeTutor in computer science education presents a nuanced perspective on their integration and effectiveness over time. Our analysis reveals a complex relationship between the duration of CodeTutor use and students' attitudes towards it. Specifically, our results show that although students initially find CodeTutor a reliable tool for understanding their queries, their confidence in its accuracy diminishes with prolonged use. Additionally, our model uncovers a weakly significant decrease in students' preference for CodeTutor as a TA replacement over time. This trend implies a growing inclination among students to seek human TA support as they progress in their courses, possibly due to the nuanced understanding and personalized feedback that human TAs can offer, which might not be fully replicated by LLMs. However, our study found no significant temporal change in students' attitudes toward CodeTutor's impact on critical thinking, syntax mastery, and independent learning. This stability suggests that while students may question CodeTutor's comprehension abilities and its adequacy as a TA replacement over time, they still recognize its utility in facilitating certain aspects of the learning process, such as mastering syntax and promoting independent study habits.

Collectively, our findings highlight the importance of investigating the temporal dynamics of student attitudes towards and their use of LLM-powered tools for learning and shed light on the need for a balanced approach to integrating LLMs into CS education. While these tools offer great support in specific areas, their limitations become more apparent with extended use. In other words, it is important to complement LLMs with human instruction to address learning objectives, such as critical thinking and problem-solving, which are crucial for computer science education. Furthermore, we argue that educators and developers should work collaboratively to enhance the capabilities of LLM-powered tutoring systems, ensuring they remain effective and relevant over time.

\subsection{Alignments of LLMs for Education} 
Our observations regarding students' utilization of CodeTutor provide insights into their learning approaches and completion of assignments. The exams that prohibit using CodeTutor reflect students' understanding of programming, as they must rely solely on their internal knowledge. Conversely, assignments and lab tasks that permit using CodeTutor result in higher scores, indicating that students may prioritize completion over deep comprehension~\cite{gustafson2022}. While students employ CodeTutor to fulfill homework requirements, they may not perceive it as a tool for a comprehensive understanding of course materials.

Our results show that nearly half of the low-quality prompts classified as \textit{over-reliance} were copied and pasted original assignment questions into CodeTutor. This suggests that students primarily used CodeTutor as a quick-fix solution, neglecting the opportunity to engage with the underlying question logic and determine appropriate solutions to the question. As the complexity of assignments increased, students' perceptions of CodeTutor's ability to understand their queries turned more negative. However, students acknowledge its proficiency in syntax mastery, which reveals a gap between their expectations and the tool's capabilities. Complex questions require students to integrate and apply the knowledge acquired in class~\cite{trautwein2003relationship}, challenging the notion that CodeTutor can easily break down questions into manageable components. Additionally, CodeTutor's limitations, such as its training on a predetermined database and inability to handle custom or complex queries, suggest that it is important to simplify questions and structure prompts effectively for optimal results.

Furthermore, we argue that students’ previous experiences with chatbots, if unrelated to structured learning, such as a simple one-line request (e.g., ``help me write a summary''), may not adequately prepare them for using CodeTutor effectively in a programming context, as evidenced by our findings that nearly 70\% of student submissions in our corpus were of poor quality. Students with limited experience interacting with chatbots might be hesitant to trust tools like CodeTutor fully, potentially affecting their use and reliance on its outputs. This lack of familiarity could lead them to prefer traditional learning approaches, fostering deeper analytical thinking and minimizing dependency on automated assistance. 

\textbf{Design Implications. } 
Our findings shed light on the future implementation and enhancement of CodeTutor in programming courses. The inherent limitations of CodeTutor, which is trained on a general dataset, may necessitate the creation of custom datasets tailored to specific class contexts. Through instructors' reflections on the quality of students' assignments, it becomes evident that while CodeTutor produces impressive results due to its training on datasets crafted by professional programmers aimed at efficiency, the emphasis in entry-level classes should prioritize human-readable code over complex solutions. One potential solution is to leverage GPT models with the Assistant API~\cite{assistantAPI}. This API enables the development of AI assistants with features, such as the Code Interpreter~\cite{code_interpreter}, which can execute Python code in a sandboxed environment, and Knowledge Retrieval~\cite{knowledge_retrieval}, allowing users to upload documents to enhance the assistant's knowledge base. These features align more closely with the requirements of a virtual TA in entry-level programming courses. The Code Interpreter can enhance the quality of responses containing code blocks, while Knowledge Retrieval empowers instructors to provide course-specific information. Meanwhile, providing systematic instructions to students can enhance their understanding of how to use the tool effectively while improving its accessibility through additional instructional features. Additionally, it is crucial to emphasize the boundaries of using LLM-powered tools, clarifying what is permissible and the consequences of inappropriate usage.

\section{Limitations and Future Work}
Our study, while providing valuable insights into the use of LLM-powered tools in educational settings, has several limitations that suggest avenues for further research. First, The current study was conducted on a relatively small scale, limiting the generalizability of our findings. Therefore, our future work will conduct larger-scale tests involving more diverse student populations and settings. Second, regarding the applicability to different levels of coding courses, our work has focused on beginning levels of CS courses. Our findings may not directly translate to intermediate or advanced programming courses. Furthermore, we relied on GPT-3.5 in this study, which may not always provide accurate or contextually appropriate responses, potentially affecting the quality of tutoring provided. Lastly, controlling the experimental environment in a semester-long study, particularly the control group, was challenging, indicating the need for more experimental designs in future studies to better understand the factors affecting student learning.

\section{Conclusion}
In this work, we conducted a semester-long between-subjects study with 50 students to examine the ways in which students use an LLM-powered virtual teaching assistant (i.e., CodeTutor) in their introductory-level programming learning. The experimental group using CodeTutor showed significant improvements in final scores over the control group, with first-time users of LLM-powered tools experiencing the most substantial gains. While positive feedback was received on CodeTutor's ability to understand queries and aid in syntax learning, concerns were raised about its effectiveness in cultivating critical thinking skills. Over time, we observed a shift towards preferring human assistant support over CodeTutor, despite its utility in completing programming tasks, understanding syntax, and debugging. Our study also shows the importance of prompt quality in leveraging CodeTutor's effectiveness, indicating that detailed and clear prompts yield more accurate responses. Our findings point to the critical need for embedding \textit{Generative AI literacy} into educational curricula and to promote critical thinking abilities among students. Looking ahead, our research suggests integrating LLM-powered tools in computer science education requires more tools, resources, and regulations to help students develop Generative AI literacy and customize teaching strategies to bridge the gap between tool capabilities and educational goals. By adjusting expectations and guiding students on effective tool use, educators may harness the full potential of Generative AI to complement traditional teaching methods.

\begin{acks}
This project is funded by the Studio for Teaching \& Learning Innovation Learn, Discover, Innovate Grant, the Faculty Research Grant from William \& Mary, and the Microsoft Accelerate Foundation Models Research Award. We thank our participants in this study and our anonymous reviewers for their feedback.
\end{acks}

\bibliographystyle{ACM-Reference-Format}
\bibliography{ref}


\begin{thebibliography}{60}


\ifx \showCODEN    \undefined \def \showCODEN     #1{\unskip}     \fi
\ifx \showDOI      \undefined \def \showDOI       #1{#1}\fi
\ifx \showISBNx    \undefined \def \showISBNx     #1{\unskip}     \fi
\ifx \showISBNxiii \undefined \def \showISBNxiii  #1{\unskip}     \fi
\ifx \showISSN     \undefined \def \showISSN      #1{\unskip}     \fi
\ifx \showLCCN     \undefined \def \showLCCN      #1{\unskip}     \fi
\ifx \shownote     \undefined \def \shownote      #1{#1}          \fi
\ifx \showarticletitle \undefined \def \showarticletitle #1{#1}   \fi
\ifx \showURL      \undefined \def \showURL       {\relax}        \fi
\providecommand\bibfield[2]{#2}
\providecommand\bibinfo[2]{#2}
\providecommand\natexlab[1]{#1}
\providecommand\showeprint[2][]{arXiv:#2}

\bibitem[Achiam et~al\mbox{.}(2023)]%
        {achiam2023gpt}
\bibfield{author}{\bibinfo{person}{Josh Achiam}, \bibinfo{person}{Steven
  Adler}, \bibinfo{person}{Sandhini Agarwal}, \bibinfo{person}{Lama Ahmad},
  \bibinfo{person}{Ilge Akkaya}, \bibinfo{person}{Florencia~Leoni Aleman},
  \bibinfo{person}{Diogo Almeida}, \bibinfo{person}{Janko Altenschmidt},
  \bibinfo{person}{Sam Altman}, \bibinfo{person}{Shyamal Anadkat},
  {et~al\mbox{.}}} \bibinfo{year}{2023}\natexlab{}.
\newblock \showarticletitle{Gpt-4 technical report}.
\newblock \bibinfo{journal}{\emph{arXiv preprint arXiv:2303.08774}}
  (\bibinfo{year}{2023}).
\newblock
\urldef\tempurl%
\url{https://doi.org/10.48550/arXiv.2303.08774}
\showDOI{\tempurl}


\bibitem[Ahmed et~al\mbox{.}(2022)]%
        {ahmed2022synshine}
\bibfield{author}{\bibinfo{person}{Toufique Ahmed}, \bibinfo{person}{Noah~Rose
  Ledesma}, {and} \bibinfo{person}{Premkumar Devanbu}.}
  \bibinfo{year}{2022}\natexlab{}.
\newblock \showarticletitle{SYNSHINE: improved fixing of syntax errors}.
\newblock \bibinfo{journal}{\emph{IEEE Transactions on Software Engineering}}
  \bibinfo{volume}{49}, \bibinfo{number}{4} (\bibinfo{year}{2022}),
  \bibinfo{pages}{2169--2181}.
\newblock
\urldef\tempurl%
\url{https://doi.org/10.1109/TSE.2022.3212635}
\showDOI{\tempurl}


\bibitem[Anderson et~al\mbox{.}(1985)]%
        {anderson1985intelligent}
\bibfield{author}{\bibinfo{person}{John~R Anderson},
  \bibinfo{person}{C~Franklin Boyle}, {and} \bibinfo{person}{Brian~J Reiser}.}
  \bibinfo{year}{1985}\natexlab{}.
\newblock \showarticletitle{Intelligent tutoring systems}.
\newblock \bibinfo{journal}{\emph{Science}} \bibinfo{volume}{228},
  \bibinfo{number}{4698} (\bibinfo{year}{1985}), \bibinfo{pages}{456--462}.
\newblock
\urldef\tempurl%
\url{https://doi.org/10.1126/science.228.4698.456}
\showDOI{\tempurl}


\bibitem[Bates et~al\mbox{.}(2015)]%
        {lme4cite}
\bibfield{author}{\bibinfo{person}{Douglas Bates}, \bibinfo{person}{Martin
  M{\"a}chler}, \bibinfo{person}{Ben Bolker}, {and} \bibinfo{person}{Steve
  Walker}.} \bibinfo{year}{2015}\natexlab{}.
\newblock \showarticletitle{Fitting Linear Mixed-Effects Models Using {lme4}}.
\newblock \bibinfo{journal}{\emph{Journal of Statistical Software}}
  \bibinfo{volume}{67}, \bibinfo{number}{1} (\bibinfo{year}{2015}),
  \bibinfo{pages}{1--48}.
\newblock
\urldef\tempurl%
\url{https://doi.org/10.18637/jss.v067.i01}
\showDOI{\tempurl}


\bibitem[Brusilovsky et~al\mbox{.}(1998)]%
        {brusilovsky1998adaptive}
\bibfield{author}{\bibinfo{person}{Peter Brusilovsky} {et~al\mbox{.}}}
  \bibinfo{year}{1998}\natexlab{}.
\newblock \showarticletitle{Adaptive educational systems on the world-wide-web:
  A review of available technologies}. In \bibinfo{booktitle}{\emph{Proceedings
  of Workshop" WWW-Based Tutoring" at 4th International Conference on
  Intelligent Tutoring Systems (ITS'98), San Antonio, TX}}.
\newblock


\bibitem[Brusilovsky et~al\mbox{.}(1996)]%
        {brusilovsky1996elm}
\bibfield{author}{\bibinfo{person}{Peter Brusilovsky}, \bibinfo{person}{Elmar
  Schwarz}, {and} \bibinfo{person}{Gerhard Weber}.}
  \bibinfo{year}{1996}\natexlab{}.
\newblock \showarticletitle{ELM-ART: An intelligent tutoring system on World
  Wide Web}. In \bibinfo{booktitle}{\emph{Intelligent Tutoring Systems: Third
  International Conference, ITS'96 Montr{\'e}al, Canada, June 12--14, 1996
  Proceedings 3}}. Springer, \bibinfo{pages}{261--269}.
\newblock
\urldef\tempurl%
\url{https://doi.org/10.1007/3-540-61327-7_123}
\showDOI{\tempurl}


\bibitem[Butz et~al\mbox{.}(2006)]%
        {butz2006web}
\bibfield{author}{\bibinfo{person}{Cory~J Butz}, \bibinfo{person}{Shan Hua},
  {and} \bibinfo{person}{R~Brien Maguire}.} \bibinfo{year}{2006}\natexlab{}.
\newblock \showarticletitle{A web-based bayesian intelligent tutoring system
  for computer programming}.
\newblock \bibinfo{journal}{\emph{Web Intelligence and Agent Systems: An
  International Journal}} \bibinfo{volume}{4}, \bibinfo{number}{1}
  (\bibinfo{year}{2006}), \bibinfo{pages}{77--97}.
\newblock


\bibitem[Clusmann et~al\mbox{.}(2023)]%
        {llm_medicine_2}
\bibfield{author}{\bibinfo{person}{Jan Clusmann}, \bibinfo{person}{Fiona~R
  Kolbinger}, \bibinfo{person}{Hannah~Sophie Muti}, \bibinfo{person}{Zunamys~I
  Carrero}, \bibinfo{person}{Jan-Niklas Eckardt},
  \bibinfo{person}{Narmin~Ghaffari Laleh}, \bibinfo{person}{Chiara
  Maria~Lavinia L{\"o}ffler}, \bibinfo{person}{Sophie-Caroline Schwarzkopf},
  \bibinfo{person}{Michaela Unger}, \bibinfo{person}{Gregory~P Veldhuizen},
  {et~al\mbox{.}}} \bibinfo{year}{2023}\natexlab{}.
\newblock \showarticletitle{The future landscape of large language models in
  medicine}.
\newblock \bibinfo{journal}{\emph{Communications Medicine}}
  \bibinfo{volume}{3}, \bibinfo{number}{1} (\bibinfo{year}{2023}),
  \bibinfo{pages}{141}.
\newblock
\urldef\tempurl%
\url{https://doi.org/10.1038/s43856-023-00370-1}
\showDOI{\tempurl}


\bibitem[Corbett et~al\mbox{.}(1997)]%
        {corbett1997intelligent}
\bibfield{author}{\bibinfo{person}{Albert~T Corbett},
  \bibinfo{person}{Kenneth~R Koedinger}, {and} \bibinfo{person}{John~R
  Anderson}.} \bibinfo{year}{1997}\natexlab{}.
\newblock \showarticletitle{Intelligent tutoring systems}.
\newblock In \bibinfo{booktitle}{\emph{Handbook of human-computer
  interaction}}. \bibinfo{publisher}{Elsevier}, \bibinfo{pages}{849--874}.
\newblock
\urldef\tempurl%
\url{https://doi.org/10.1016/B978-044481862-1.50103-5}
\showDOI{\tempurl}


\bibitem[Demszky and Liu(2023)]%
        {demszky2023m}
\bibfield{author}{\bibinfo{person}{Dorottya Demszky} {and}
  \bibinfo{person}{Jing Liu}.} \bibinfo{year}{2023}\natexlab{}.
\newblock \showarticletitle{M-Powering Teachers: Natural Language Processing
  Powered Feedback Improves 1: 1 Instruction and Student Outcomes}.
\newblock  (\bibinfo{year}{2023}).
\newblock
\urldef\tempurl%
\url{https://doi.org/10.1145/3573051.3593379}
\showDOI{\tempurl}


\bibitem[Denny et~al\mbox{.}(2022)]%
        {denny2022robosourcing}
\bibfield{author}{\bibinfo{person}{Paul Denny}, \bibinfo{person}{Sami Sarsa},
  \bibinfo{person}{Arto Hellas}, {and} \bibinfo{person}{Juho Leinonen}.}
  \bibinfo{year}{2022}\natexlab{}.
\newblock \showarticletitle{Robosourcing Educational Resources--Leveraging
  Large Language Models for Learnersourcing}.
\newblock \bibinfo{journal}{\emph{arXiv preprint arXiv:2211.04715}}
  (\bibinfo{year}{2022}).
\newblock
\urldef\tempurl%
\url{https://doi.org/10.1145/3501385.3543957}
\showDOI{\tempurl}


\bibitem[Dobslaw and Bergh(2023)]%
        {dobslaw2023experiences}
\bibfield{author}{\bibinfo{person}{Felix Dobslaw} {and} \bibinfo{person}{Peter
  Bergh}.} \bibinfo{year}{2023}\natexlab{}.
\newblock \showarticletitle{Experiences with Remote Examination Formats in
  Light of GPT-4}.
\newblock \bibinfo{journal}{\emph{arXiv preprint arXiv:2305.02198}}
  (\bibinfo{year}{2023}).
\newblock
\urldef\tempurl%
\url{https://doi.org/10.48550/arXiv.2305.02198}
\showDOI{\tempurl}


\bibitem[El~Saadawi et~al\mbox{.}(2008)]%
        {el2008natural}
\bibfield{author}{\bibinfo{person}{Gilan~M El~Saadawi}, \bibinfo{person}{Eugene
  Tseytlin}, \bibinfo{person}{Elizabeth Legowski}, \bibinfo{person}{Drazen
  Jukic}, \bibinfo{person}{Melissa Castine}, \bibinfo{person}{Jeffrey Fine},
  \bibinfo{person}{Robert Gormley}, {and} \bibinfo{person}{Rebecca~S Crowley}.}
  \bibinfo{year}{2008}\natexlab{}.
\newblock \showarticletitle{A natural language intelligent tutoring system for
  training pathologists: Implementation and evaluation}.
\newblock \bibinfo{journal}{\emph{Advances in health sciences education}}
  \bibinfo{volume}{13} (\bibinfo{year}{2008}), \bibinfo{pages}{709--722}.
\newblock
\urldef\tempurl%
\url{https://doi.org/10.1007/s10459-007-9081-3}
\showDOI{\tempurl}


\bibitem[Elsom-Cook(1984)]%
        {elsom1984design}
\bibfield{author}{\bibinfo{person}{Mark Elsom-Cook}.}
  \bibinfo{year}{1984}\natexlab{}.
\newblock \emph{\bibinfo{title}{Design considerations of an intelligent
  tutoring system for programming languages}}.
\newblock \bibinfo{thesistype}{Ph.\,D. Dissertation}.
  \bibinfo{school}{University of Warwick}.
\newblock


\bibitem[{GitHub, Inc.}(2024)]%
        {GitHubCopilot}
\bibfield{author}{\bibinfo{person}{{GitHub, Inc.}}}
  \bibinfo{year}{2024}\natexlab{}.
\newblock \bibinfo{title}{GitHub Copilot}.
\newblock \bibinfo{howpublished}{\url{https://github.com/features/copilot}}.
\newblock
\newblock
\shownote{Accessed: 2024-02-11}.


\bibitem[Graesser et~al\mbox{.}(2018)]%
        {graesser2018intelligent}
\bibfield{author}{\bibinfo{person}{Arthur~C Graesser}, \bibinfo{person}{Xiangen
  Hu}, {and} \bibinfo{person}{Robert Sottilare}.}
  \bibinfo{year}{2018}\natexlab{}.
\newblock \showarticletitle{Intelligent tutoring systems}.
\newblock In \bibinfo{booktitle}{\emph{International handbook of the learning
  sciences}}. \bibinfo{publisher}{Routledge}, \bibinfo{pages}{246--255}.
\newblock


\bibitem[Gustafson(2022)]%
        {gustafson2022}
\bibfield{author}{\bibinfo{person}{Morgan Gustafson}.}
  \bibinfo{year}{2022}\natexlab{}.
\newblock \bibinfo{title}{The Effect of Homework Completion on Students'
  Academic Performance}.
\newblock \bibinfo{howpublished}{Dissertations, Theses, and Projects}.
\newblock
\urldef\tempurl%
\url{https://red.mnstate.edu/thesis/662}
\showURL{%
\tempurl}
\newblock
\shownote{662}.


\bibitem[Hicke et~al\mbox{.}(2023)]%
        {hicke2023chata}
\bibfield{author}{\bibinfo{person}{Yann Hicke}, \bibinfo{person}{Anmol
  Agarwal}, \bibinfo{person}{Qianou Ma}, {and} \bibinfo{person}{Paul Denny}.}
  \bibinfo{year}{2023}\natexlab{}.
\newblock \showarticletitle{ChaTA: Towards an Intelligent Question-Answer
  Teaching Assistant using Open-Source LLMs}.
\newblock \bibinfo{journal}{\emph{arXiv preprint arXiv:2311.02775}}
  (\bibinfo{year}{2023}).
\newblock
\urldef\tempurl%
\url{https://doi.org/10.48550/arXiv.2311.02775}
\showDOI{\tempurl}


\bibitem[Hooshyar et~al\mbox{.}(2015)]%
        {hooshyar2015flowchart}
\bibfield{author}{\bibinfo{person}{Danial Hooshyar},
  \bibinfo{person}{Rodina~Binti Ahmad}, \bibinfo{person}{Moslem Yousefi},
  \bibinfo{person}{Farrah~Dina Yusop}, {and} \bibinfo{person}{S-J Horng}.}
  \bibinfo{year}{2015}\natexlab{}.
\newblock \showarticletitle{A flowchart-based intelligent tutoring system for
  improving problem-solving skills of novice programmers}.
\newblock \bibinfo{journal}{\emph{Journal of computer assisted learning}}
  \bibinfo{volume}{31}, \bibinfo{number}{4} (\bibinfo{year}{2015}),
  \bibinfo{pages}{345--361}.
\newblock
\urldef\tempurl%
\url{https://doi.org/10.1111/jcal.12099}
\showDOI{\tempurl}


\bibitem[Jalil et~al\mbox{.}(2023)]%
        {jalil2023chatgpt}
\bibfield{author}{\bibinfo{person}{Sajed Jalil}, \bibinfo{person}{Suzzana
  Rafi}, \bibinfo{person}{Thomas~D LaToza}, \bibinfo{person}{Kevin Moran},
  {and} \bibinfo{person}{Wing Lam}.} \bibinfo{year}{2023}\natexlab{}.
\newblock \showarticletitle{Chatgpt and software testing education: Promises \&
  perils}. In \bibinfo{booktitle}{\emph{2023 IEEE International Conference on
  Software Testing, Verification and Validation Workshops (ICSTW)}}. IEEE,
  \bibinfo{pages}{4130--4137}.
\newblock
\urldef\tempurl%
\url{https://doi.org/10.1109/ICSTW58534.2023.00078}
\showDOI{\tempurl}


\bibitem[Kasneci et~al\mbox{.}(2023)]%
        {kasneci2023chatgpt}
\bibfield{author}{\bibinfo{person}{Enkelejda Kasneci}, \bibinfo{person}{Kathrin
  Se{\ss}ler}, \bibinfo{person}{Stefan K{\"u}chemann}, \bibinfo{person}{Maria
  Bannert}, \bibinfo{person}{Daryna Dementieva}, \bibinfo{person}{Frank
  Fischer}, \bibinfo{person}{Urs Gasser}, \bibinfo{person}{Georg Groh},
  \bibinfo{person}{Stephan G{\"u}nnemann}, \bibinfo{person}{Eyke
  H{\"u}llermeier}, {et~al\mbox{.}}} \bibinfo{year}{2023}\natexlab{}.
\newblock \showarticletitle{ChatGPT for good? On opportunities and challenges
  of large language models for education}.
\newblock \bibinfo{journal}{\emph{Learning and individual differences}}
  \bibinfo{volume}{103} (\bibinfo{year}{2023}), \bibinfo{pages}{102274}.
\newblock
\urldef\tempurl%
\url{https://doi.org/10.1016/j.lindif.2023.102274}
\showDOI{\tempurl}


\bibitem[Kulik and Fletcher(2016)]%
        {kulik2016effectiveness}
\bibfield{author}{\bibinfo{person}{James~A Kulik} {and} \bibinfo{person}{JD
  Fletcher}.} \bibinfo{year}{2016}\natexlab{}.
\newblock \showarticletitle{Effectiveness of intelligent tutoring systems: a
  meta-analytic review}.
\newblock \bibinfo{journal}{\emph{Review of educational research}}
  \bibinfo{volume}{86}, \bibinfo{number}{1} (\bibinfo{year}{2016}),
  \bibinfo{pages}{42--78}.
\newblock
\urldef\tempurl%
\url{https://doi.org/10.3102/0034654315581420}
\showDOI{\tempurl}


\bibitem[Kumar et~al\mbox{.}(2023)]%
        {kumar2023impact}
\bibfield{author}{\bibinfo{person}{Harsh Kumar}, \bibinfo{person}{Ilya
  Musabirov}, \bibinfo{person}{Mohi Reza}, \bibinfo{person}{Jiakai Shi},
  \bibinfo{person}{Anastasia Kuzminykh}, \bibinfo{person}{Joseph~Jay Williams},
  {and} \bibinfo{person}{Michael Liut}.} \bibinfo{year}{2023}\natexlab{}.
\newblock \showarticletitle{Impact of Guidance and Interaction Strategies for
  LLM Use on Learner Performance and Perception}.
\newblock \bibinfo{journal}{\emph{arXiv preprint arXiv:2310.13712}}
  (\bibinfo{year}{2023}).
\newblock
\urldef\tempurl%
\url{https://doi.org/10.48550/arXiv.2310.13712}
\showDOI{\tempurl}


\bibitem[Leinonen et~al\mbox{.}(2023a)]%
        {leinonen2023comparing}
\bibfield{author}{\bibinfo{person}{Juho Leinonen}, \bibinfo{person}{Paul
  Denny}, \bibinfo{person}{Stephen MacNeil}, \bibinfo{person}{Sami Sarsa},
  \bibinfo{person}{Seth Bernstein}, \bibinfo{person}{Joanne Kim},
  \bibinfo{person}{Andrew Tran}, {and} \bibinfo{person}{Arto Hellas}.}
  \bibinfo{year}{2023}\natexlab{a}.
\newblock \showarticletitle{Comparing code explanations created by students and
  large language models}.
\newblock \bibinfo{journal}{\emph{arXiv preprint arXiv:2304.03938}}
  (\bibinfo{year}{2023}).
\newblock
\urldef\tempurl%
\url{https://doi.org/10.48550/arXiv.2304.03938}
\showDOI{\tempurl}


\bibitem[Leinonen et~al\mbox{.}(2023b)]%
        {leinonen2023using}
\bibfield{author}{\bibinfo{person}{Juho Leinonen}, \bibinfo{person}{Arto
  Hellas}, \bibinfo{person}{Sami Sarsa}, \bibinfo{person}{Brent Reeves},
  \bibinfo{person}{Paul Denny}, \bibinfo{person}{James Prather}, {and}
  \bibinfo{person}{Brett~A Becker}.} \bibinfo{year}{2023}\natexlab{b}.
\newblock \showarticletitle{Using large language models to enhance programming
  error messages}. In \bibinfo{booktitle}{\emph{Proceedings of the 54th ACM
  Technical Symposium on Computer Science Education V. 1}}.
  \bibinfo{pages}{563--569}.
\newblock
\urldef\tempurl%
\url{https://doi.org/10.1145/3545945.3569770}
\showDOI{\tempurl}


\bibitem[Liffiton et~al\mbox{.}({[n.\,d.]})]%
        {liffiton2023codehelp}
\bibfield{author}{\bibinfo{person}{Mark Liffiton}, \bibinfo{person}{Brad~E
  Sheese}, \bibinfo{person}{Jaromir Savelka}, {and} \bibinfo{person}{Paul
  Denny}.} \bibinfo{year}{[n.\,d.]}\natexlab{}.
\newblock \showarticletitle{Codehelp: Using large language models with
  guardrails for scalable support in programming classes}.
\newblock  (\bibinfo{year}{[n.\,d.]}), \bibinfo{pages}{1--11}.
\newblock
\urldef\tempurl%
\url{https://doi.org/10.1145/3631802.3631830}
\showDOI{\tempurl}


\bibitem[Mehta et~al\mbox{.}(2023)]%
        {mehta2023can}
\bibfield{author}{\bibinfo{person}{Atharva Mehta}, \bibinfo{person}{Nipun
  Gupta}, \bibinfo{person}{Dhruv Kumar}, \bibinfo{person}{Pankaj Jalote},
  {et~al\mbox{.}}} \bibinfo{year}{2023}\natexlab{}.
\newblock \showarticletitle{Can ChatGPT Play the Role of a Teaching Assistant
  in an Introductory Programming Course?}
\newblock \bibinfo{journal}{\emph{arXiv preprint arXiv:2312.07343}}
  (\bibinfo{year}{2023}).
\newblock
\urldef\tempurl%
\url{https://doi.org/10.48550/arXiv.2312.07343}
\showDOI{\tempurl}


\bibitem[Meyer et~al\mbox{.}(2023)]%
        {meyer2023chatgpt}
\bibfield{author}{\bibinfo{person}{Jesse~G Meyer}, \bibinfo{person}{Ryan~J
  Urbanowicz}, \bibinfo{person}{Patrick~CN Martin}, \bibinfo{person}{Karen
  O’Connor}, \bibinfo{person}{Ruowang Li}, \bibinfo{person}{Pei-Chen Peng},
  \bibinfo{person}{Tiffani~J Bright}, \bibinfo{person}{Nicholas Tatonetti},
  \bibinfo{person}{Kyoung~Jae Won}, \bibinfo{person}{Graciela
  Gonzalez-Hernandez}, {et~al\mbox{.}}} \bibinfo{year}{2023}\natexlab{}.
\newblock \showarticletitle{ChatGPT and large language models in academia:
  opportunities and challenges}.
\newblock \bibinfo{journal}{\emph{BioData Mining}} \bibinfo{volume}{16},
  \bibinfo{number}{1} (\bibinfo{year}{2023}), \bibinfo{pages}{20}.
\newblock
\urldef\tempurl%
\url{https://doi.org/10.1186/s13040-023-00339-9}
\showDOI{\tempurl}


\bibitem[Nwana(1990)]%
        {nwana1990intelligent}
\bibfield{author}{\bibinfo{person}{Hyacinth~S Nwana}.}
  \bibinfo{year}{1990}\natexlab{}.
\newblock \showarticletitle{Intelligent tutoring systems: an overview}.
\newblock \bibinfo{journal}{\emph{Artificial Intelligence Review}}
  \bibinfo{volume}{4}, \bibinfo{number}{4} (\bibinfo{year}{1990}),
  \bibinfo{pages}{251--277}.
\newblock
\urldef\tempurl%
\url{https://doi.org/10.1007/BF00168958}
\showDOI{\tempurl}


\bibitem[Ogle et~al\mbox{.}(2023)]%
        {fsacite}
\bibfield{author}{\bibinfo{person}{Derek~H. Ogle}, \bibinfo{person}{Jason~C.
  Doll}, \bibinfo{person}{A.~Powell Wheeler}, {and} \bibinfo{person}{Alexis
  Dinno}.} \bibinfo{year}{2023}\natexlab{}.
\newblock \bibinfo{booktitle}{\emph{FSA: Simple Fisheries Stock Assessment
  Methods}}.
\newblock
\urldef\tempurl%
\url{https://CRAN.R-project.org/package=FSA}
\showURL{%
\tempurl}
\newblock
\shownote{R package version 0.9.4}.


\bibitem[OpenAI(2024)]%
        {assistantAPI}
\bibfield{author}{\bibinfo{person}{OpenAI}.} \bibinfo{year}{2024}\natexlab{}.
\newblock \bibinfo{title}{Assistants Overview - OpenAI API}.
\newblock
  \bibinfo{howpublished}{\url{https://platform.openai.com/docs/assistants/overview}}.
\newblock
\newblock
\shownote{Accessed: 2024-02-11}.


\bibitem[{OpenAI}(2024)]%
        {ChatGPT}
\bibfield{author}{\bibinfo{person}{{OpenAI}}.} \bibinfo{year}{2024}\natexlab{}.
\newblock \bibinfo{title}{ChatGPT}.
\newblock \bibinfo{howpublished}{\url{https://openai.com/chatgpt}}.
\newblock
\newblock
\shownote{Accessed: 2024-02-11}.


\bibitem[OpenAI(2024a)]%
        {code_interpreter}
\bibfield{author}{\bibinfo{person}{OpenAI}.} \bibinfo{year}{2024}\natexlab{a}.
\newblock \bibinfo{title}{Code Interpreter}.
\newblock
  \bibinfo{howpublished}{\url{https://platform.openai.com/docs/assistants/tools/code-interpreter}}.
\newblock
\newblock
\shownote{Accessed: 2024-02-11}.


\bibitem[OpenAI(2024b)]%
        {knowledge_retrieval}
\bibfield{author}{\bibinfo{person}{OpenAI}.} \bibinfo{year}{2024}\natexlab{b}.
\newblock \bibinfo{title}{Knowledge Retrieval}.
\newblock
  \bibinfo{howpublished}{\url{https://platform.openai.com/docs/assistants/tools/knowledge-retrieval}}.
\newblock
\newblock
\shownote{Accessed: 2024-02-11}.


\bibitem[Pankiewicz and Baker(2023)]%
        {pankiewicz2023large}
\bibfield{author}{\bibinfo{person}{Maciej Pankiewicz} {and}
  \bibinfo{person}{Ryan~S Baker}.} \bibinfo{year}{2023}\natexlab{}.
\newblock \showarticletitle{Large Language Models (GPT) for automating feedback
  on programming assignments}.
\newblock \bibinfo{journal}{\emph{arXiv preprint arXiv:2307.00150}}
  (\bibinfo{year}{2023}).
\newblock
\urldef\tempurl%
\url{https://doi.org/10.48550/arXiv.2307.00150}
\showDOI{\tempurl}


\bibitem[Perkins et~al\mbox{.}(2023)]%
        {perkins2023detection}
\bibfield{author}{\bibinfo{person}{Mike Perkins}, \bibinfo{person}{Jasper Roe},
  \bibinfo{person}{Darius Postma}, \bibinfo{person}{James McGaughran}, {and}
  \bibinfo{person}{Don Hickerson}.} \bibinfo{year}{2023}\natexlab{}.
\newblock \showarticletitle{Detection of GPT-4 generated text in higher
  education: Combining academic judgement and software to identify generative
  AI tool misuse}.
\newblock \bibinfo{journal}{\emph{Journal of Academic Ethics}}
  (\bibinfo{year}{2023}), \bibinfo{pages}{1--25}.
\newblock
\urldef\tempurl%
\url{https://doi.org/10.1007/s10805-023-09492-6}
\showDOI{\tempurl}


\bibitem[Phung et~al\mbox{.}(2023a)]%
        {phung2023generating}
\bibfield{author}{\bibinfo{person}{Tung Phung}, \bibinfo{person}{Jos{\'e}
  Cambronero}, \bibinfo{person}{Sumit Gulwani}, \bibinfo{person}{Tobias Kohn},
  \bibinfo{person}{Rupak Majumdar}, \bibinfo{person}{Adish Singla}, {and}
  \bibinfo{person}{Gustavo Soares}.} \bibinfo{year}{2023}\natexlab{a}.
\newblock \showarticletitle{Generating High-Precision Feedback for Programming
  Syntax Errors using Large Language Models}.
\newblock \bibinfo{journal}{\emph{arXiv preprint arXiv:2302.04662}}
  (\bibinfo{year}{2023}).
\newblock
\urldef\tempurl%
\url{https://doi.org/10.48550/arXiv.2302.04662}
\showDOI{\tempurl}


\bibitem[Phung et~al\mbox{.}(2023b)]%
        {phung2023generative}
\bibfield{author}{\bibinfo{person}{Tung Phung},
  \bibinfo{person}{Victor-Alexandru P{\u{a}}durean}, \bibinfo{person}{Jos{\'e}
  Cambronero}, \bibinfo{person}{Sumit Gulwani}, \bibinfo{person}{Tobias Kohn},
  \bibinfo{person}{Rupak Majumdar}, \bibinfo{person}{Adish Singla}, {and}
  \bibinfo{person}{Gustavo Soares}.} \bibinfo{year}{2023}\natexlab{b}.
\newblock \showarticletitle{Generative AI for Programming Education:
  Benchmarking ChatGPT, GPT-4, and Human Tutors}.
\newblock \bibinfo{journal}{\emph{International Journal of Management}}
  \bibinfo{volume}{21}, \bibinfo{number}{2} (\bibinfo{year}{2023}),
  \bibinfo{pages}{100790}.
\newblock
\urldef\tempurl%
\url{https://doi.org/10.48550/arXiv.2306.17156}
\showDOI{\tempurl}


\bibitem[Poldrack et~al\mbox{.}(2023)]%
        {poldrack2023ai}
\bibfield{author}{\bibinfo{person}{Russell~A Poldrack}, \bibinfo{person}{Thomas
  Lu}, {and} \bibinfo{person}{Ga{\v{s}}per Begu{\v{s}}}.}
  \bibinfo{year}{2023}\natexlab{}.
\newblock \showarticletitle{AI-assisted coding: Experiments with GPT-4}.
\newblock \bibinfo{journal}{\emph{arXiv preprint arXiv:2304.13187}}
  (\bibinfo{year}{2023}).
\newblock
\urldef\tempurl%
\url{https://doi.org/10.48550/arXiv.2304.13187}
\showDOI{\tempurl}


\bibitem[Prather et~al\mbox{.}(2023)]%
        {prather2023robots}
\bibfield{author}{\bibinfo{person}{James Prather}, \bibinfo{person}{Paul
  Denny}, \bibinfo{person}{Juho Leinonen}, \bibinfo{person}{Brett~A Becker},
  \bibinfo{person}{Ibrahim Albluwi}, \bibinfo{person}{Michelle Craig},
  \bibinfo{person}{Hieke Keuning}, \bibinfo{person}{Natalie Kiesler},
  \bibinfo{person}{Tobias Kohn}, \bibinfo{person}{Andrew Luxton-Reilly},
  {et~al\mbox{.}}} \bibinfo{year}{2023}\natexlab{}.
\newblock \showarticletitle{The robots are here: Navigating the generative ai
  revolution in computing education}.
\newblock \bibinfo{journal}{\emph{arXiv preprint arXiv:2310.00658}}
  (\bibinfo{year}{2023}).
\newblock
\urldef\tempurl%
\url{https://doi.org/10.1145/3623762.3633499}
\showDOI{\tempurl}


\bibitem[{R Core Team}(2022)]%
        {statscite}
\bibfield{author}{\bibinfo{person}{{R Core Team}}.}
  \bibinfo{year}{2022}\natexlab{}.
\newblock \bibinfo{booktitle}{\emph{R: A Language and Environment for
  Statistical Computing}}.
\newblock R Foundation for Statistical Computing, Vienna, Austria.
\newblock
\urldef\tempurl%
\url{https://www.R-project.org/}
\showURL{%
\tempurl}


\bibitem[Ritter et~al\mbox{.}(2007)]%
        {ritter2007cognitive}
\bibfield{author}{\bibinfo{person}{Steven Ritter}, \bibinfo{person}{John~R
  Anderson}, \bibinfo{person}{Kenneth~R Koedinger}, {and}
  \bibinfo{person}{Albert Corbett}.} \bibinfo{year}{2007}\natexlab{}.
\newblock \showarticletitle{Cognitive Tutor: Applied research in mathematics
  education}.
\newblock \bibinfo{journal}{\emph{Psychonomic bulletin \& review}}
  \bibinfo{volume}{14} (\bibinfo{year}{2007}), \bibinfo{pages}{249--255}.
\newblock
\urldef\tempurl%
\url{https://doi.org/10.3758/BF03194060}
\showDOI{\tempurl}


\bibitem[Sarsa et~al\mbox{.}(2022)]%
        {sarsa2022automatic}
\bibfield{author}{\bibinfo{person}{Sami Sarsa}, \bibinfo{person}{Paul Denny},
  \bibinfo{person}{Arto Hellas}, {and} \bibinfo{person}{Juho Leinonen}.}
  \bibinfo{year}{2022}\natexlab{}.
\newblock \showarticletitle{Automatic generation of programming exercises and
  code explanations using large language models}. In
  \bibinfo{booktitle}{\emph{Proceedings of the 2022 ACM Conference on
  International Computing Education Research-Volume 1}}.
  \bibinfo{pages}{27--43}.
\newblock


\bibitem[Savelka et~al\mbox{.}(2023)]%
        {savelka2023large}
\bibfield{author}{\bibinfo{person}{Jaromir Savelka}, \bibinfo{person}{Arav
  Agarwal}, \bibinfo{person}{Christopher Bogart}, {and} \bibinfo{person}{Majd
  Sakr}.} \bibinfo{year}{2023}\natexlab{}.
\newblock \showarticletitle{Large language models (gpt) struggle to answer
  multiple-choice questions about code}.
\newblock \bibinfo{journal}{\emph{arXiv preprint arXiv:2303.08033}}
  (\bibinfo{year}{2023}).
\newblock
\urldef\tempurl%
\url{https://doi.org/10.48550/arXiv.2303.08033}
\showDOI{\tempurl}


\bibitem[Sheese et~al\mbox{.}(2023)]%
        {sheese2023patterns}
\bibfield{author}{\bibinfo{person}{Brad Sheese}, \bibinfo{person}{Mark
  Liffiton}, \bibinfo{person}{Jaromir Savelka}, {and} \bibinfo{person}{Paul
  Denny}.} \bibinfo{year}{2023}\natexlab{}.
\newblock \showarticletitle{Patterns of Student Help-Seeking When Using a Large
  Language Model-Powered Programming Assistant}.
\newblock \bibinfo{journal}{\emph{arXiv preprint arXiv:2310.16984}}
  (\bibinfo{year}{2023}).
\newblock
\urldef\tempurl%
\url{https://doi.org/10.1145/3636243.3636249}
\showDOI{\tempurl}


\bibitem[Sleeman and Brown(1982)]%
        {sleeman1982intelligent}
\bibfield{author}{\bibinfo{person}{Derek Sleeman} {and}
  \bibinfo{person}{John~Seely Brown}.} \bibinfo{year}{1982}\natexlab{}.
\newblock \bibinfo{booktitle}{\emph{Intelligent tutoring systems}}.
\newblock \bibinfo{publisher}{London: Academic Press}.
\newblock


\bibitem[Sottilare et~al\mbox{.}(2012)]%
        {sottilare2012generalized}
\bibfield{author}{\bibinfo{person}{Robert~A Sottilare},
  \bibinfo{person}{Keith~W Brawner}, \bibinfo{person}{Benjamin~S Goldberg},
  {and} \bibinfo{person}{Heather~K Holden}.} \bibinfo{year}{2012}\natexlab{}.
\newblock \showarticletitle{The generalized intelligent framework for tutoring
  (GIFT)}.
\newblock \bibinfo{journal}{\emph{Orlando, FL: US Army Research
  Laboratory--Human Research \& Engineering Directorate (ARL-HRED)}}
  (\bibinfo{year}{2012}).
\newblock


\bibitem[Sun et~al\mbox{.}(2024)]%
        {sun2024trustllm}
\bibfield{author}{\bibinfo{person}{Lichao Sun}, \bibinfo{person}{Yue Huang},
  \bibinfo{person}{Haoran Wang}, \bibinfo{person}{Siyuan Wu},
  \bibinfo{person}{Qihui Zhang}, \bibinfo{person}{Chujie Gao},
  \bibinfo{person}{Yixin Huang}, \bibinfo{person}{Wenhan Lyu},
  \bibinfo{person}{Yixuan Zhang}, \bibinfo{person}{Xiner Li}, {et~al\mbox{.}}}
  \bibinfo{year}{2024}\natexlab{}.
\newblock \showarticletitle{Trustllm: Trustworthiness in large language
  models}.
\newblock \bibinfo{journal}{\emph{arXiv preprint arXiv:2401.05561}}
  (\bibinfo{year}{2024}).
\newblock
\urldef\tempurl%
\url{https://doi.org/10.48550/arXiv.2401.05561}
\showDOI{\tempurl}


\bibitem[Thirunavukarasu et~al\mbox{.}(2023)]%
        {llm_medicine_1}
\bibfield{author}{\bibinfo{person}{Arun~James Thirunavukarasu},
  \bibinfo{person}{Darren Shu~Jeng Ting}, \bibinfo{person}{Kabilan Elangovan},
  \bibinfo{person}{Laura Gutierrez}, \bibinfo{person}{Ting~Fang Tan}, {and}
  \bibinfo{person}{Daniel Shu~Wei Ting}.} \bibinfo{year}{2023}\natexlab{}.
\newblock \showarticletitle{Large language models in medicine}.
\newblock \bibinfo{journal}{\emph{Nature medicine}} \bibinfo{volume}{29},
  \bibinfo{number}{8} (\bibinfo{year}{2023}), \bibinfo{pages}{1930--1940}.
\newblock
\urldef\tempurl%
\url{https://doi.org/10.1038/s41591-023-02448-8}
\showDOI{\tempurl}


\bibitem[Thomas(2006)]%
        {thomas2006general}
\bibfield{author}{\bibinfo{person}{David~R Thomas}.}
  \bibinfo{year}{2006}\natexlab{}.
\newblock \showarticletitle{A general inductive approach for analyzing
  qualitative evaluation data}.
\newblock \bibinfo{journal}{\emph{American journal of evaluation}}
  \bibinfo{volume}{27}, \bibinfo{number}{2} (\bibinfo{year}{2006}),
  \bibinfo{pages}{237--246}.
\newblock
\urldef\tempurl%
\url{https://doi.org/10.1177/1098214005283748}
\showDOI{\tempurl}


\bibitem[Trautwein and K{\"o}ller(2003)]%
        {trautwein2003relationship}
\bibfield{author}{\bibinfo{person}{Ulrich Trautwein} {and}
  \bibinfo{person}{Olaf K{\"o}ller}.} \bibinfo{year}{2003}\natexlab{}.
\newblock \showarticletitle{The relationship between homework and
  achievement—still much of a mystery}.
\newblock \bibinfo{journal}{\emph{Educational psychology review}}
  \bibinfo{volume}{15} (\bibinfo{year}{2003}), \bibinfo{pages}{115--145}.
\newblock
\urldef\tempurl%
\url{https://doi.org/10.1023/A:1023460414243}
\showDOI{\tempurl}


\bibitem[VanLehn(2011)]%
        {vanlehn2011relative}
\bibfield{author}{\bibinfo{person}{Kurt VanLehn}.}
  \bibinfo{year}{2011}\natexlab{}.
\newblock \showarticletitle{The relative effectiveness of human tutoring,
  intelligent tutoring systems, and other tutoring systems}.
\newblock \bibinfo{journal}{\emph{Educational psychologist}}
  \bibinfo{volume}{46}, \bibinfo{number}{4} (\bibinfo{year}{2011}),
  \bibinfo{pages}{197--221}.
\newblock
\urldef\tempurl%
\url{https://doi.org/10.1080/00461520.2011.611369}
\showDOI{\tempurl}


\bibitem[Wei et~al\mbox{.}(2022)]%
        {wei2022emergent}
\bibfield{author}{\bibinfo{person}{Jason Wei}, \bibinfo{person}{Yi Tay},
  \bibinfo{person}{Rishi Bommasani}, \bibinfo{person}{Colin Raffel},
  \bibinfo{person}{Barret Zoph}, \bibinfo{person}{Sebastian Borgeaud},
  \bibinfo{person}{Dani Yogatama}, \bibinfo{person}{Maarten Bosma},
  \bibinfo{person}{Denny Zhou}, \bibinfo{person}{Donald Metzler},
  {et~al\mbox{.}}} \bibinfo{year}{2022}\natexlab{}.
\newblock \showarticletitle{Emergent abilities of large language models}.
\newblock \bibinfo{journal}{\emph{arXiv preprint arXiv:2206.07682}}
  (\bibinfo{year}{2022}).
\newblock
\urldef\tempurl%
\url{https://doi.org/10.48550/arXiv.2206.07682}
\showDOI{\tempurl}


\bibitem[Wu et~al\mbox{.}(2023b)]%
        {wu2023survey}
\bibfield{author}{\bibinfo{person}{Junchao Wu}, \bibinfo{person}{Shu Yang},
  \bibinfo{person}{Runzhe Zhan}, \bibinfo{person}{Yulin Yuan},
  \bibinfo{person}{Derek~F Wong}, {and} \bibinfo{person}{Lidia~S Chao}.}
  \bibinfo{year}{2023}\natexlab{b}.
\newblock \showarticletitle{A survey on llm-gernerated text detection:
  Necessity, methods, and future directions}.
\newblock \bibinfo{journal}{\emph{arXiv preprint arXiv:2310.14724}}
  (\bibinfo{year}{2023}).
\newblock
\urldef\tempurl%
\url{https://doi.org/10.48550/arXiv.2310.14724}
\showDOI{\tempurl}


\bibitem[Wu et~al\mbox{.}(2023a)]%
        {llm_finance_1}
\bibfield{author}{\bibinfo{person}{Shijie Wu}, \bibinfo{person}{Ozan Irsoy},
  \bibinfo{person}{Steven Lu}, \bibinfo{person}{Vadim Dabravolski},
  \bibinfo{person}{Mark Dredze}, \bibinfo{person}{Sebastian Gehrmann},
  \bibinfo{person}{Prabhanjan Kambadur}, \bibinfo{person}{David Rosenberg},
  {and} \bibinfo{person}{Gideon Mann}.} \bibinfo{year}{2023}\natexlab{a}.
\newblock \showarticletitle{Bloomberggpt: A large language model for finance}.
\newblock \bibinfo{journal}{\emph{arXiv preprint arXiv:2303.17564}}
  (\bibinfo{year}{2023}).
\newblock
\urldef\tempurl%
\url{https://doi.org/10.48550/arXiv.2303.17564}
\showDOI{\tempurl}


\bibitem[Zamfirescu-Pereira et~al\mbox{.}(2023)]%
        {zamfirescu2023johnny}
\bibfield{author}{\bibinfo{person}{JD Zamfirescu-Pereira},
  \bibinfo{person}{Richmond~Y Wong}, \bibinfo{person}{Bjoern Hartmann}, {and}
  \bibinfo{person}{Qian Yang}.} \bibinfo{year}{2023}\natexlab{}.
\newblock \showarticletitle{Why Johnny can’t prompt: how non-AI experts try
  (and fail) to design LLM prompts}. In \bibinfo{booktitle}{\emph{Proceedings
  of the 2023 CHI Conference on Human Factors in Computing Systems}}.
  \bibinfo{pages}{1--21}.
\newblock
\urldef\tempurl%
\url{https://doi.org/10.1145/3544548.3581388}
\showDOI{\tempurl}


\bibitem[Zhang et~al\mbox{.}(2024)]%
        {zhang2024profiling}
\bibfield{author}{\bibinfo{person}{Yixuan Zhang}, \bibinfo{person}{Yimeng
  Wang}, \bibinfo{person}{Nutchanon Yongsatianchot}, \bibinfo{person}{Joseph~D
  Gaggiano}, \bibinfo{person}{Nurul~M Suhaimi}, \bibinfo{person}{Anne Okrah},
  \bibinfo{person}{Miso Kim}, \bibinfo{person}{Jacqueline Griffin}, {and}
  \bibinfo{person}{Andrea~G Parker}.} \bibinfo{year}{2024}\natexlab{}.
\newblock \showarticletitle{Profiling the Dynamics of Trust \& Distrust in
  Social Media: A Survey Study}.
\newblock  (\bibinfo{year}{2024}).
\newblock
\urldef\tempurl%
\url{https://doi.org/10.1145/3613904.3642927}
\showDOI{\tempurl}


\bibitem[Zhou et~al\mbox{.}(2023)]%
        {zhou2023synthetic}
\bibfield{author}{\bibinfo{person}{Jiawei Zhou}, \bibinfo{person}{Yixuan
  Zhang}, \bibinfo{person}{Qianni Luo}, \bibinfo{person}{Andrea~G Parker},
  {and} \bibinfo{person}{Munmun De~Choudhury}.}
  \bibinfo{year}{2023}\natexlab{}.
\newblock \showarticletitle{Synthetic lies: Understanding ai-generated
  misinformation and evaluating algorithmic and human solutions}. In
  \bibinfo{booktitle}{\emph{Proceedings of the 2023 CHI Conference on Human
  Factors in Computing Systems}}. \bibinfo{pages}{1--20}.
\newblock
\urldef\tempurl%
\url{https://doi.org/10.1145/3544548.3581318}
\showDOI{\tempurl}


\bibitem[Zhou et~al\mbox{.}(2024)]%
        {zhou2024teachers}
\bibfield{author}{\bibinfo{person}{Kyrie~Zhixuan Zhou},
  \bibinfo{person}{Zachary Kilhoffer}, \bibinfo{person}{Madelyn~Rose
  Sanfilippo}, \bibinfo{person}{Ted Underwood}, \bibinfo{person}{Ece Gumusel},
  \bibinfo{person}{Mengyi Wei}, \bibinfo{person}{Abhinav Choudhry}, {and}
  \bibinfo{person}{Jinjun Xiong}.} \bibinfo{year}{2024}\natexlab{}.
\newblock \showarticletitle{"The teachers are confused as well": A
  Multiple-Stakeholder Ethics Discussion on Large Language Models in Computing
  Education}.
\newblock \bibinfo{journal}{\emph{arXiv preprint arXiv:2401.12453}}
  (\bibinfo{year}{2024}).
\newblock
\urldef\tempurl%
\url{https://doi.org/10.48550/arXiv.2401.12453}
\showDOI{\tempurl}


\bibitem[Zhou et~al\mbox{.}(2022)]%
        {zhou2022large}
\bibfield{author}{\bibinfo{person}{Yongchao Zhou}, \bibinfo{person}{Andrei~Ioan
  Muresanu}, \bibinfo{person}{Ziwen Han}, \bibinfo{person}{Keiran Paster},
  \bibinfo{person}{Silviu Pitis}, \bibinfo{person}{Harris Chan}, {and}
  \bibinfo{person}{Jimmy Ba}.} \bibinfo{year}{2022}\natexlab{}.
\newblock \showarticletitle{Large language models are human-level prompt
  engineers}.
\newblock \bibinfo{journal}{\emph{arXiv preprint arXiv:2211.01910}}
  (\bibinfo{year}{2022}).
\newblock
\urldef\tempurl%
\url{https://doi.org/10.48550/arXiv.2211.01910}
\showDOI{\tempurl}


\end{thebibliography}


\end{document}